\documentclass[preprint2]{aastex6}
\fullcollaborationName{The Friends of AASTeX Collaboration}
\begin{document}


\title{Dust concentration and  emission in protoplanetary disks vortices}


\author{Anibal Sierra \altaffilmark{1}, Susana Lizano \altaffilmark{1}, and Pierre Barge\altaffilmark{2}}


\altaffiltext{1}{Instituto de Radioastronom\'ia y Astrof\'isica, UNAM, Apartado Postal 3-72, 58089 Morelia Michoac\'an, M\'exico}
\altaffiltext{2}{Aix Marseille Universit\'e, CNRS, Laboratoire d'Astrophysique de Marseille, UMR 7326, 13388, Marseille, France}

\begin{abstract}

We study the dust concentration and emission in protoplanetary disks vortices.  We extend the Lyra-Lin solution for the dust concentration of a single grain size to a power-law distribution of grain sizes $n(a) \propto a^{-p}$.
Assuming dust conservation in the disk, we find an analytic dust surface density as a function of the grain radius. We calculate the increase of the dust to gas mass ratio $\epsilon$ and the slope $p$ of the dust size distribution due to grain segregation within the vortex.
We apply this model to a numerical simulation of a disk containing a persistent vortex.
Due to the accumulation of large grains towards the vortex center, 
$\epsilon$ increases by a factor of 10 from the background disk value, and $p$ decreases from 3.5 to 3.0. We find the disk emission at millimeter wavelengths corresponding to synthetic observations with ALMA and VLA. The simulated maps at 7 mm and 1 cm show a strong azimuthal asymmetry. This happens because, at these wavelengths,  the disk becomes optically thin while the vortex remains optically thick. 
The large vortex opacity is mainly due to an increase in the dust to gas mass ratio. In addition, the change in the slope of the dust size distribution increases the opacity by a factor of 2. We also show that the inclusion of the dust scattering opacity substantially changes the disks images. 

\end{abstract}




\section{Introduction} \label{sec:intro}

There is not yet a full theory that successfully explains the formation of solid planet that starts with the concentration and growth of dust particles in protoplanetary disks from millimeter to planetesimal sizes. One of the main problems is the fast radial migration of millimeter and micrometer dust particles toward the central star, which prevents the formation of large bodies during the disk lifetime (e.g., \citealt{Testi_2014}; \citealt{Johansen_2014}). This inward drift is due to the collisions of the dust grains (which tend to rotate at the Keplerian speed) with the gas molecules (which flow at sub-Keplerian speed), causing the loss of the dust angular momentum and, thus, dust radial migration. A natural way to prevent the fast migration is growth, because the inward radial velocity is a function of the dust grain size. For example, in the minimum mass solar nebula (MMSN) model, the inward radial migration has a maximum
speed  for 1 m objects \citep{Weidenschilling_1977}. Thus, the dust particles should grow to sizes larger than 1 m in order to prevent their fast radial migration. However, to build meter-sized bodies (or larger) via dust collisions is not very effective, because the typical collision velocities are so violent  that the final result is dust fragmentation instead of coagulation \citep{Brauer_2008}. These two obstacles are known as the radial drift barrier and the fragmentation barrier, respectively.

One of the ideas proposed in order to avoid this problem is dust trapping in pressure bumps. \cite{Barge_1995} found that persistent gaseous vortices can effectively concentrate and segregate large amounts of solid particles via pressure gradients, possibly starting the formation of planetesimals. The dynamics of large scale vortices has been studied using 2D \citep{Surville_2015} and 3D \citep{Richard_2013} hydrodynamical simulations in protoplanetary disks; these azimuthal asymmetries naturally arise due to the Rossby wave instability \citep{Li_2000} or the baroclinic instability (e.g. \citealt{Barge_2016}) in the outer edge of the dead zone, where turbulence due to the Magneto Rotational Instability (MRI) is  depressed due to the low ionization state of the disk material. These structures can survive over a hundred rotation periods (measured at the radius of the center of the vortex) and increase the dust to gas mass ratio one order of magnitude \citep{Inaba_2006}. This large concentration of dust mass could become gravitationally unstable, and start the formation of planetesimals.

In the last years,  high angular resolution mm observations with ALMA and VLA, and infrared observations with SPHERE on the VLT have found large scale structures in several sources that could be the signatures of vortices in the disk. Some examples are the disks around the young stars Oph IRS 48 \citep{VanderMarel_2013}, HD 142527 \citep{Casassus_2013}, LkH$\alpha$ \citep{Isella_2013},  MWC 758 \citep{Marino_2015}, SAO 206462, and SR 21 \citep{Perez_2014}.  Spiral arms structures have also been found in the dust emission, e.g., around MWC 758 and Elias 2-24 (\cite{Benisty_2015}; \cite{Perez_2016}).
These observations are important to  understand the physical processes suffered by the dust during the formation of planetesimals and the gas dynamics in protoplanetary disks;
they also set important constrains to the parameters used in the theoretical models.

In this paper we consider the millimeter emission of a dust vortex in a protoplanetary disk obtained from the gas numerical simulation performed by \cite{Barge_2017} (hereafter B17). We extend the Lyra-Lin dust vortex model \citep{Lyra_2013} for a single grain size to a power-law distribution of grain sizes. 
To calculate the dust millimeter emission we compute dust opacities in different regions of the vortex according to the size segregation, which changes the dust to gas mass ratio and the slope in the particle size distribution. We include the dust opacity due to both scattering and true absorption.
The dust surface density model (Section \S \ref{sec:models}) allows us to calculate the local dust properties (particle size distribution and dust to gas mass ratio) as a function of the position in the disk vortex . The Section \S \ref{sec:emission} presents the simulated disk observations at mm wavelengths and their Spectral Energy Distribution (SED).
In the Section \S \ref{sec:discussion} we compare with the emission of a model without dust segregation (\S \ref{subsec:dust_distribution}), and without scattering (\S \ref{subsec:absorption}). In subsection \S  \ref{subsec:dust_to_gas}, we discuss the maximum dust to gas mass ratio obtained in the vortex with the expected values in numerical simulations. And finally, in \S \ref{subsec:max_grain_radius}, we discuss the case when the maximum grain radius in the dust particle size distribution is $1$ cm. The conclusions are presented in the Section \S \ref{sec:conclusions}.

\section{Disk model} \label{sec:models}
\subsection{Gas Disk Model} \label{subsec:data}
\begin{figure*}[!t]
\centering
\includegraphics[scale=0.4]{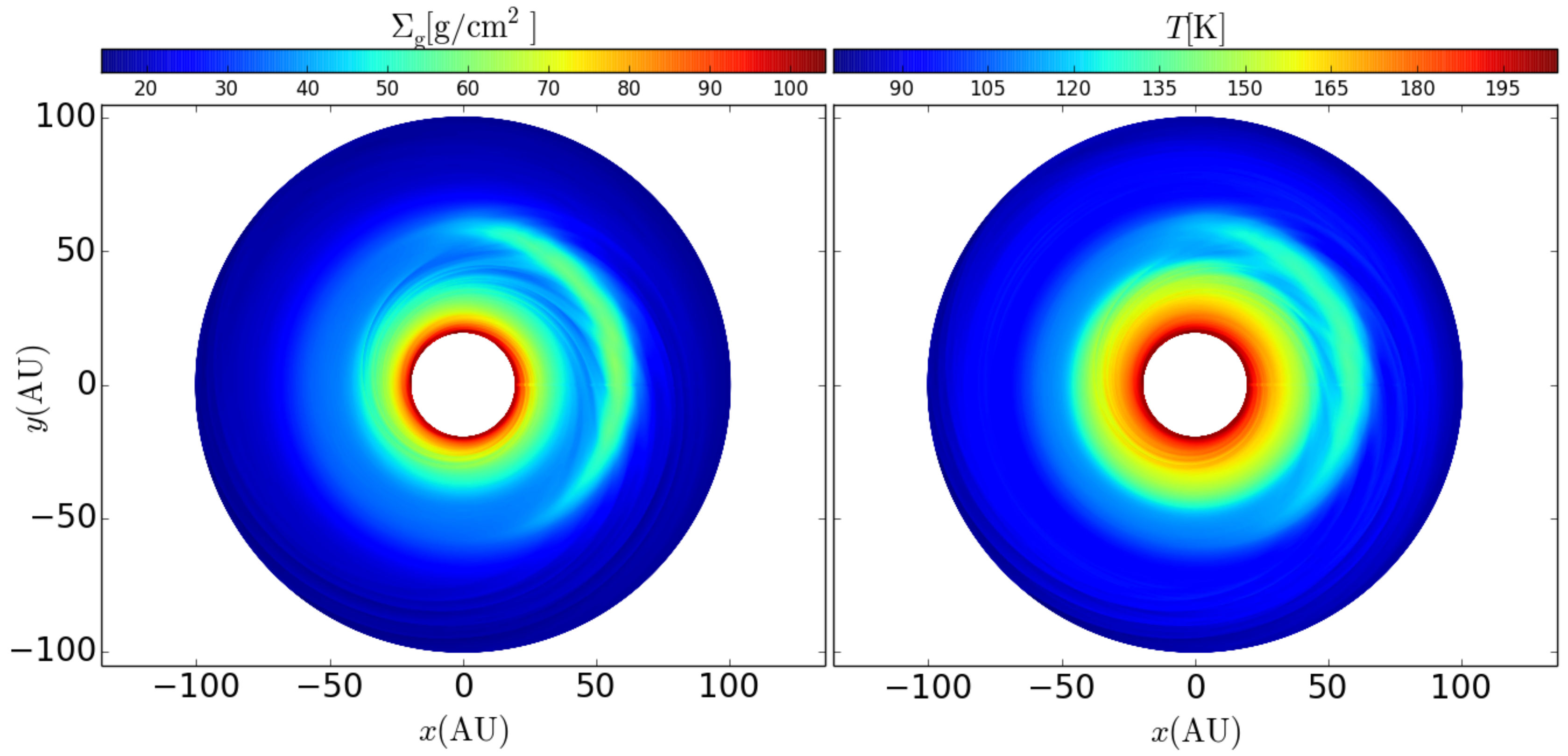}
\caption{Gas surface density (left) and gas temperature (right) of a disk containing a large scale vortex (after 100 vortex rotations) .}
\label{fig:gas_data}
\end{figure*}

For the  disk and the vortex, we use the result of numerical simulations of B17 performed over 100 rotations of the vortex around a $2 M_{\odot}$ star. 
The vortex is located at 60 AU from the star. The disk has a mass $M_{\rm d} = 0.14 M_\odot$ 
and a disk radius $R_{\rm d} = 100$ AU, which mimic the disk around the Oph IRS 48 young star. 
Figure \ref{fig:gas_data} shows the gas surface density $\Sigma_g$ and the gas temperature $T$  of the disk as a function of the cartesian coordinates ($x, y$).  
In the vertical direction $(z)$, the disk is assumed to be isothermal and in hydrostatic equilibrium. We assume a distance $d = 120$ pc, similar to the distance to Oph IRS 48.
\newpage
\subsection{Dust Disk Model} \label{subsec:dust_disk_model}
In a gaseous vortex, the dust particles tend to drift toward the pressure maximum at a rate that depends on their coupling with the gas, measured by the Stokes number 
\begin{equation}
{St} = 
\frac{\pi}{2} \frac{\rho_m a}{\Sigma_g},
\label{eq:stokes_number}
\end{equation}
where $a$ is the radius of spherical dust particles, $\rho_m = 3$ g cm$^{-3}$  is the material density and $\Sigma_g$ is the gas surface density. 
\citet{Lyra_2013} found an analytic formulation for the dust concentration in a vortex as a function of the normalized Stokes number ($S = St/\alpha$), where $\alpha$ is the turbulent viscosity parameter in the vortex \citep{Shakura_1973}. In their model, the surface density of the gas and dust depends on the semi minor axis coordinate $b$ that defines concentric ellipses with an aspect ratio $\chi$, such that the semi major axis is $\chi b$. For each ellipse, the gas and dust surface density are given by
\begin{equation}
\Sigma_{\rm g,V}(b) = \Sigma_{\rm g,max} \exp \left( - \frac{b^2}{2H_{\rm v}^2} \right),
\label{Eq:Gas_density_vortex}
\end{equation}
\begin{equation}
\Sigma_{\rm d,V}(b,a)= \frac{\Sigma_{\rm d,max}(a)}{\sqrt{S_{\rm v}+1}} \exp \left( -\frac{b^2}{2H_{\rm v}^2} (S_{\rm v}+1) \right),
\label{eq:Dust_surface_density}
\end{equation}
where $\Sigma_{\rm g,max}$, $\Sigma_{\rm d,max}(a)$ are the gas and dust maximum surface densities, $H_{\rm v} = H/f$ is the vortex scale length, where $f < 1$ multiplies the isothermal scale height $H$, this factor depends on the aspect ratio of the vortex $\chi$. The normalized Stokes number
\begin{equation}
S_{\rm v} = \frac{\pi \rho_m a}{2 \alpha \Sigma_{\rm g,max}},
\end{equation}
is evaluated in the vortex center, since the vortex is small compared with the disk size, and the gas surface density does not vary much over its area.

For a given position in the vortex with polar coordinates $(\varpi, \theta)$, the corresponding ellipse has a coordinate
\begin{equation}
b = \sqrt{\left( \varpi - \varpi_0 \right)^2  + \frac{\varpi_0^2}{\chi^2} \left( \theta - \theta_0 \right)^2},
\label{Eq:b}
\end{equation}
where $(\theta_0, \varpi_0)$ are the coordinates of the vortex center.

The total gas surface density  and the dust surface density of grains with radius $a$ can be written as
\begin{eqnarray}
\Sigma_{\rm g} &=& \Sigma_{\rm g, V}(b) + \Sigma_{\rm g,back}(\varpi), \label{Eq:Gas_density}\\
\Sigma_{\rm d}(a) &=& \Sigma_{\rm d,V}(b,a) + \Sigma_{\rm d,back}(\varpi,a),
\label{Eq:Dust_density}
\end{eqnarray} 
where $\Sigma_{\rm g,back}(\varpi), \Sigma_{\rm d,back}(\varpi,a)$ are the background disk gas and dust surface densities, respectively.
 
Since $\Sigma_{\rm d}(a)$ is a function of the grain size, the total surface density of the dust depends on the particle size distribution, $n(a)da \propto a^{-p}da$, that gives the number of dust particles per unit volume with a radius between $a$ and $a+da$. A typical value for the slope in protoplanetary disk is $p=3.5$ \citep{Mathis_1977}.\\
Integrating the eqs. (\ref{Eq:Gas_density}, \ref{Eq:Dust_density}) in all the disk, the mass of gas and the mass of dust with grain radius $a$ are

\begin{eqnarray}
M_{\rm g} &=& 2\pi \chi H_{\rm v}^2  \Sigma_{\rm g,max} + M_{\rm g,back}, \\
M_{\rm d}(a) &=& \frac{2\pi \chi H_{\rm v}^2 \Sigma_{\rm d,max}(a)}{\left( S_{\rm v}+1 \right)^{3/2}}+ M_{\rm d,back}(a),
\end{eqnarray}
where $M_{\rm g,back}$ is the gas mass of the background disk and $M_{\rm d,back}(a)$ is the background disk
mass of dust with radius $a$.

We define the global dust to gas mass ratio $\epsilon(a)$ of particles with size $a$ as

\begin{equation}
\epsilon(a) \equiv  \frac{M_{\rm d}(a)}{M_{\rm g}} = \frac{\frac{2\pi \chi H_{\rm v}^2 \Sigma_{\rm d,max}(a)}{\left( S_{\rm v}+1 \right)^{3/2}}+ M_{\rm d,back}(a)}{2\pi \chi H_{\rm v}^2  \Sigma_{\rm g,max} + M_{\rm g,back}}.
\end{equation}

We assume that $\epsilon(a)$  is conserved in the disk, i.e. the background disk also satisfies $\epsilon(a) = M_{\rm d,back}(a) /M_{\rm g,back}$. This implies that the dust is redistributed in the vortex but does not coagulate and/or fragment.
Therefore, the dust maximum surface density for a size $a$ is
\begin{equation}
\Sigma_{\rm d,max}(a) = \epsilon(a) \Sigma_{\rm g,max} \left( S_{\rm v}+1 \right)^{3/2}.
\label{Eq:Dust_density_max}
\end{equation}

For a particle size distribution in the background disk with $p=3.5$, the mass is dominated by the large grains. If $M_{d}^c(a)$ is the cumulative dust mass from the minimum grain size $a_{\rm min}$ to a size $a$, then, the ratio between the cumulative mass and the total dust mass in the disk ($M_{\rm d}^{T}$) is

\begin{equation}
\frac{M_{\rm d}^{c}(a)}{M_{\rm d}^{T}} = \frac{\int _{a_{\rm min}} ^{a} a^3 a^{-3.5} da }{\int _{a_{\rm min}} ^{a_{\rm max}} a^3 a^{-3.5} da } \approx \sqrt{\frac{a}{a_{\rm max}}},
 \label{Eq:Size_Dis}
\end{equation}
where we have assumed that $a_{\rm max} >> a_{\rm min}$. Furthermore, because the dust mass is dominated by the large dust particles, we approximate the dust mass of the grain population with size $a$ by the cumulative mass, i.e.,  $M_{\rm d}(a) \sim M_{\rm d}^{c}(a) $. Thus, equation (\ref{Eq:Size_Dis}) can be rewritten as
\begin{equation}
\frac{\epsilon(a)}{\epsilon} = \sqrt{\frac{a}{a_{\rm max}}},
\label{Eq:D2G_size}
\end{equation}
where $\epsilon = M_{\rm d}^T/M_{\rm g}$ is the total dust to gas mass ratio taking into account all the dust grain sizes.

Then, using the eqs. (\ref{Eq:Dust_density}, \ref{Eq:Dust_density_max}, \ref{Eq:D2G_size}), the dust surface density for grains with size $a$ is
\begin{eqnarray}
\nonumber \Sigma_{\rm d} (a)&=& \epsilon \sqrt{\frac{a}{a_{\rm max}}} \left[ \Sigma_{\rm g,max} (S_{\rm v}+1) \exp \left( -\frac{b^2}{2H_{\rm v}^2} (S_{\rm v}+1) \right)  \right.\\
 & &  + \Sigma_{\rm g,back}(\varpi) \bigg].
\label{Eq:Dust_Density_Fin}
\end{eqnarray}
In this equation, we have written the background dust surface density as $\Sigma_{\rm d,back}(\varpi,a) = \epsilon(a) \Sigma_{\rm g,back}(\varpi)$.

For the hydrodynamic simulation of B17, the center of the vortex is located at $\theta_0 = 0.37$ rad and $\varpi_0 = 59.16$ AU. We fit the gas surface density by using a function of the form
\begin{equation}
\Sigma_{\rm g} = A_1 \exp \left( - \frac{b^2}{2\sigma_b^2} \right) + A_2 \left(\frac{\varpi}{60 \mathrm{AU}} \right)^{-q},
\end{equation}
where the first term represents the vortex structure (see eqs. \ref{Eq:Gas_density_vortex}, \ref{Eq:b}) and the second term is the background surface density. The best fit of the numerical data leads to: $A_1 = 30.34$ g cm$^{-2}$, $\sigma_b = 5.14$ AU, $\chi = 9.4$, $A_2 = 28.9$ g cm$^{-2}$, $q = 1.1$. Note that $A_1 = \Sigma_{g,{\rm max}}$ and $\sigma_b = H_{\rm v}$ in eq. (\ref{Eq:Gas_density_vortex}).

For a disk in hydrostatic equilibrium, the dust particle size distribution is related with the dust surface density (see eq. \ref{eq:dust_distribution_sigma}), as
\begin{equation}
n(a)= c \left[ a^{-3} \frac{d \Sigma_{{d}}(a)}{da}  \right],
\end{equation}
where $c= {3}/{( 2^{5/2} \pi^{3/2} H_d \rho_m)}$ and $H_d$ is the dust scale height.
Then, from eq. (\ref{Eq:Dust_Density_Fin}), the dust particle size distribution is
\begin{eqnarray}
\nonumber \frac{n(a)}{\epsilon  c} & =  &  \left( \frac{a^{-3.5}}{a_{\rm max}^{1/2}} \right)
\left \{  
\Sigma_{\rm g,max} \left(S_{\rm v}+1 \right) \exp \left( -\frac{b^2}{2H_{\rm v}^2} (S_{\rm v}+1) \right) \right .\\
& & \times \left .  \left[  \frac{S_{\rm v}}{S_{\rm v}+1} - \frac{b^2S_{\rm v}}{2H_{\rm v}^2}  + \frac{1}{2} \right]  +  \frac{\Sigma_{\rm g,back}(\varpi)}{2} \right\} . 
\label{Eq:Dust_particle_size_distribution}
\end{eqnarray}

\begin{figure*}[!t]
\includegraphics[scale=0.35]{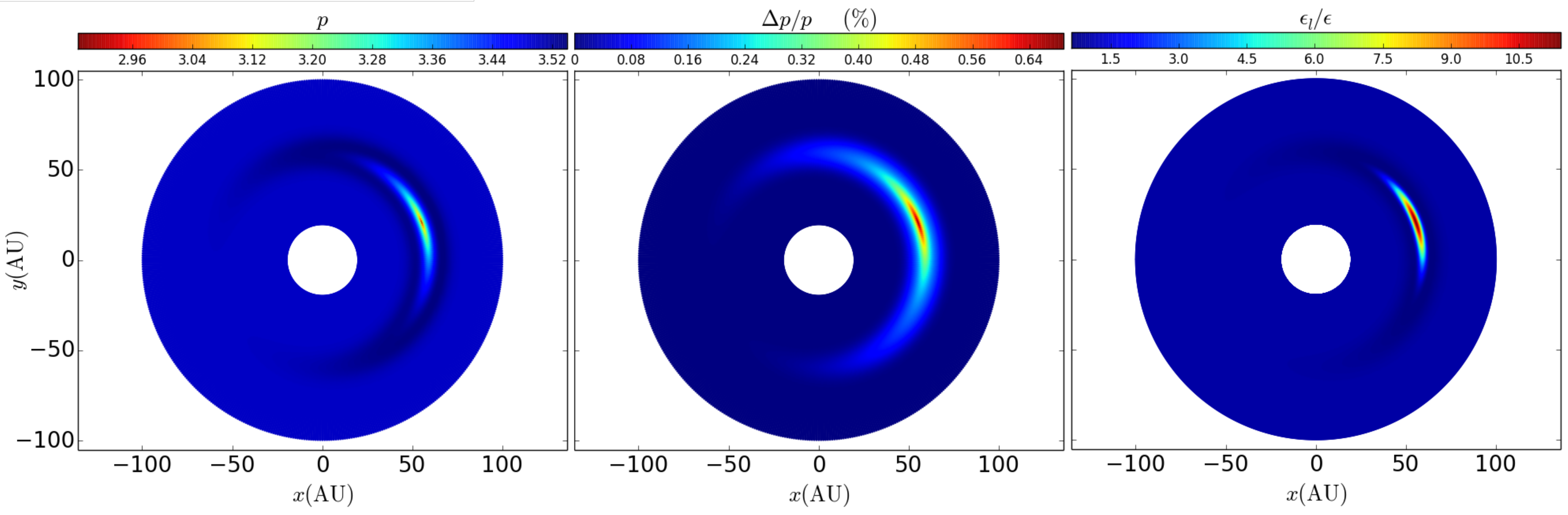}
\caption{Dust particle accumulation within the vortex. Left panel: Slope $p$ of the power law fit to the dust particle size distribution $n(a)da \propto a^{-p}da$ (see text). Middle panel: fractional standard deviation} $\Delta p/p$. Right panel: Map of the local dust to gas mass ratio (normalized to $\epsilon$).
\label{fig:dust_dist_fit_p}
\end{figure*}

The first term of this equation represents the concentration of the dust particles within the vortex, it is only important for $b \lesssim H_{\rm v}$.
We approximate eq. (\ref{Eq:Dust_particle_size_distribution}) with a simple power law function of the form $\log n(a) = -p \log(a) +k$. The left and middle panels of Figure \ref{fig:dust_dist_fit_p} show the slope $p$, the fractional standard deviation ($\Delta p/p$). The slope $p$ decreases in the vortex and reaches a minimum value $p \sim 3.0$ at the vortex center. The fit is very good as shown by the low values of the  fractional standard deviation.

Since the dust mass is dominated by the large grains, the local dust to gas mass ratio ($\epsilon_{l}$) reduces to
\begin{equation}
\epsilon_{l} \approx \frac{\Sigma_{\rm d}(a_{\rm max})}{\Sigma_{\rm g}}.
\end{equation}

This function is shown in the right panel of Figure \ref{fig:dust_dist_fit_p}, normalized to the total dust to gas mass ratio ($\epsilon$). The maximum dust to gas mass ratio ($\epsilon_{l, \rm max}$) is quite large at the vortex center, reaching a value $\epsilon_{l, \rm max} /\epsilon \sim 10.5$. Note that the increase of the dust to gas mass ratio in the vortex center is due to the accumulation of dust grains from a small region around the vortex, while the rest of the disk remains with the standard $\epsilon$ value.


\section{Dust emission} \label{sec:emission}
Now we solve the radiative transfer equation to obtain the SED and the dust emission maps of the face-on disk at different wavelengths.
The disk images are convolved with ALMA and VLA beams to simulate high angular resolution observations that are able to reveal the vortex structure. 

\subsection{Methodology} \label{subsec:methodology}
We use a grid of $1040 \times 1040$ pixels in the plane of the sky and solve the radiative transfer equation along the line of sight in order to obtain the emergent specific intensity $I_{\nu}$,

\begin{equation}
\frac{dI_{\nu}}{d\tau_{\nu}} = - I_{\nu} + S_{\nu}.
\end{equation}
The dust optical depth along the line of sight is given by
\begin{equation}
d\tau_{\nu} = \chi_\nu \rho_g dZ,
\label{eq:opacity}
\end{equation}
where $Z$ is the coordinate in the line of sight, $\rho_g$ is the volumetric gas density, and 
the total monochromatic mass opacity is $\chi_{\nu} = \kappa_{\nu} + \sigma_{\nu}$, where
the scattering and absorption mass opacity coefficients are 
$\sigma_{\nu}$ and $ \kappa_{\nu}$, respectively. The source function is given by \citep{Mihalas_1978}

\begin{equation}
S_{\nu} =  \omega_{\nu} J_{\nu} + (1- \omega_{\nu}) B_{\nu}(T),
\label{eq:source_function}
\end{equation}
where the albedo is $\omega_{\nu} = \sigma_{\nu}/(\kappa_{\nu} + \sigma_{\nu})$,  $B_{\nu}(T)$ is the Planck function,  and $J_{\nu}$  is the zeroth order moment of the specific intensity, $J_{\nu} =  \frac{1}{2} \int_{-1}^{+1} I_{\nu} d\mu$, where $\mu$ is the cosine of the angle between the direction perpendicular to the disk plane and the direction of $I_{\nu}$. We use the analytical solution of $J_{\nu}$ for a plane-parallel isotropically scattering medium found by \cite{Miyake_1993} 
 
\begin{equation}
\frac{J_{\nu} (\tau_\nu)}{B_{\nu}(T)}  = 1 + \frac{e^{-\sqrt{3\epsilon_{\nu}} \tau_\nu} + e^{ \sqrt{3\epsilon_{\nu}} (\tau_\nu - \tau_{\nu}^{d})} }{ e^{-\sqrt{3\epsilon_{\nu}} \tau_{\nu}^{d}} (\sqrt{\epsilon_{\nu}} - 1 )  - (\sqrt{\epsilon_{\nu}} +1) } ,
\label{eq:zero_moment}
\end{equation}
where $\epsilon_{\nu} = 1 - \omega_{\nu}$, and $\tau_{\nu}^{d}$ is the total optical depth of the disk measured perpendicular to plane of the disk.

The total monochromatic opacity $\chi_{\nu}$ is computed with the code from \cite{Dalessio_2001}. For the dust composition, we adopt a mixture of silicates, organics and ice with a mass fractional abundance relative to the gas $\epsilon_{\rm sil} = 3.4 \times 10^{-3}$, $\epsilon_{\rm org} = 4.1 \times 10^{-3}$, and $\epsilon_{\rm ice} = 5.6 \times 10^{-3}$. This implies a total dust to gas mass ratio $\epsilon = 0.0131$. The material densities are $\rho_{\rm sil} = 3.3$ g cm$^{-3}$, $\rho_{\rm org}= 1.5$ g cm$^{-3}$, and $\rho_{\rm ice} = 0.92$ g cm$^{-3}$ (e.g. \citealt{Pollack_1994}). We assume a dust particle size distribution $n(a)da \propto a^{-p}da$, where $p$ varies according to the disk region; the minimum and maximum dust radii are  $a_{\rm min} = 0.05\, \mu$m and $a_{\rm max}=1$ mm. The maximum grain radius is  larger in protoplanetary disks than in the ISM due to grain growth. Grain growth occurs due to coagulation during collisions induced by Brownian motions (important only for $\mu$m particles), radial drift, vertical settling and turbulent mixing; it is, however, counteracted by fragmentation and bouncing of the mm-cm particles, which collide at high velocities  (e.g. \citealt{Brauer_2008}; \citealt{Zsom_2010}).
Recent radio observations of protoplanetary disks also infer a radial dust size gradient with dust grains of centimeter sizes in the inner disk regions and millimeter sizes in the outer disk regions (e.g. \citealt{Perez_2015}). 

\subsection{Images and SED} \label{subsec:images}
We use CASA (v 4.7.0) \footnote{CASA, the Common Astronomy Software Applications package, is a software developed to support data processing of radio astronomical telescopes.} to simulate ALMA observations of the disk model shown in Figure \ref{fig:gas_data} with the dust properties of Figure \ref{fig:dust_dist_fit_p}. We calculate the images at 1 and 3 mm for the configurations C4-6 (Band 7) and C4-8 (Band 4). 
The precipitable water vapor (PWV) was set to 1.3 and 1.5 mm in the Band 7 and Band 4 configuration, respectively. These antenna arrays provide a similar angular resolution $\theta_B \sim 0\farcs10$, which corresponds to 12 AU at the assumed distance of 120 pc. 

We also simulate VLA observations at 7 mm and 1 cm with CASA in the A configuration  (the highest resolution). The FWHM beam of these configurations are $\theta_B \sim 0\farcs043$ and $0\farcs19$, respectively. In both simulations the PWV was set to 1.5 mm.

\begin{figure*}[!t]
\centering
\includegraphics[scale=0.5]{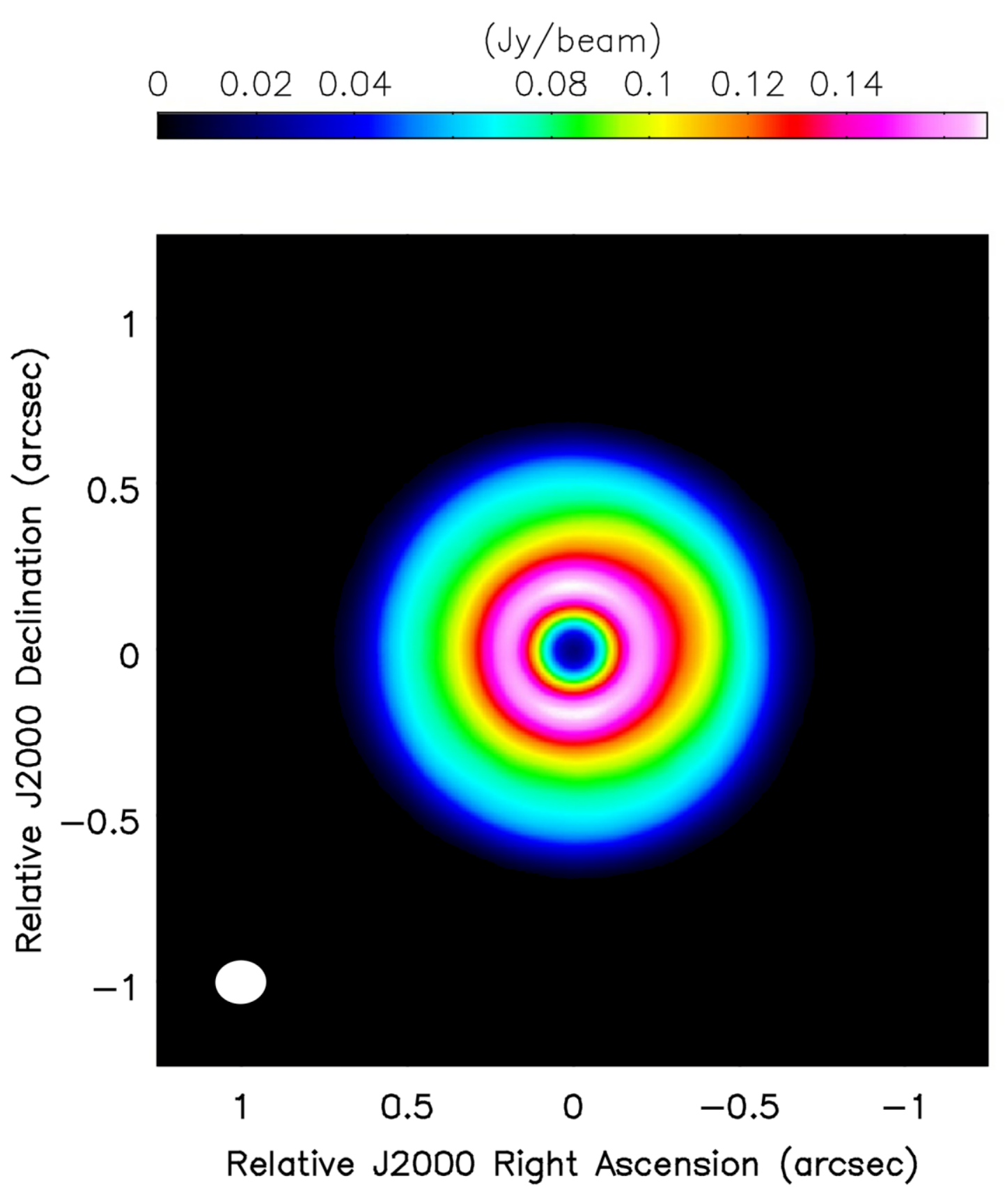}
\includegraphics[scale=0.5]{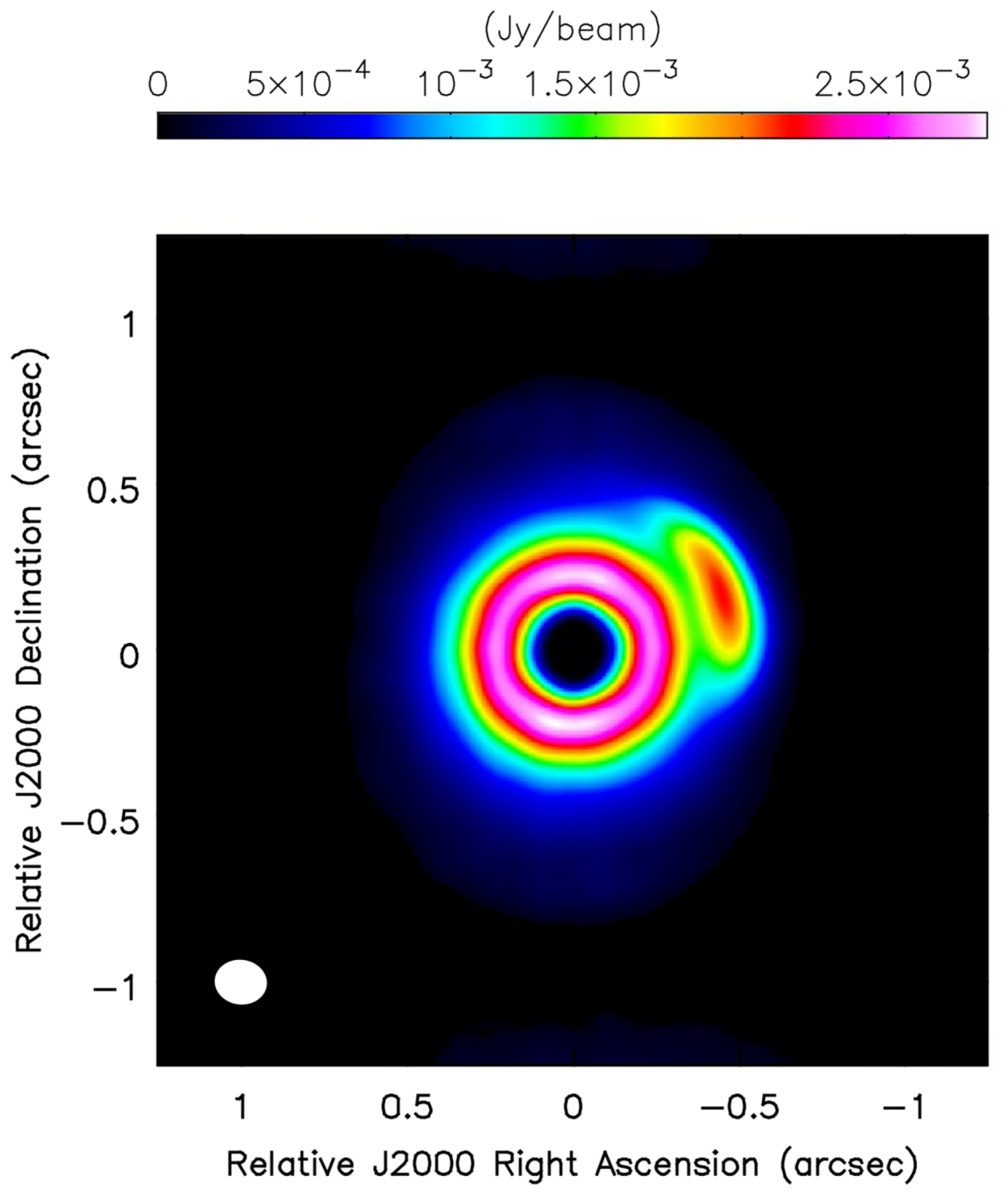}
\includegraphics[scale=0.5]{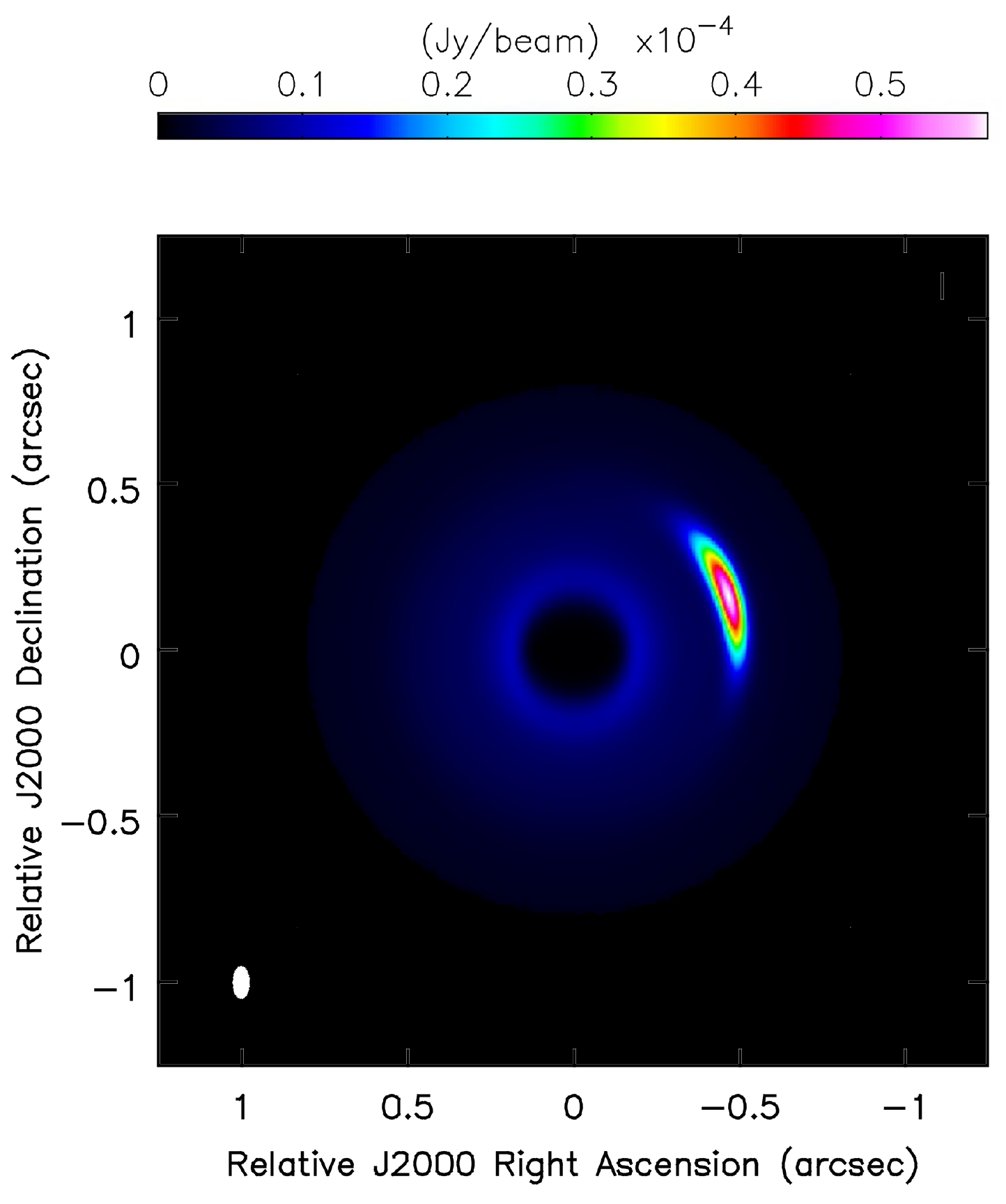}
\includegraphics[scale=0.5]{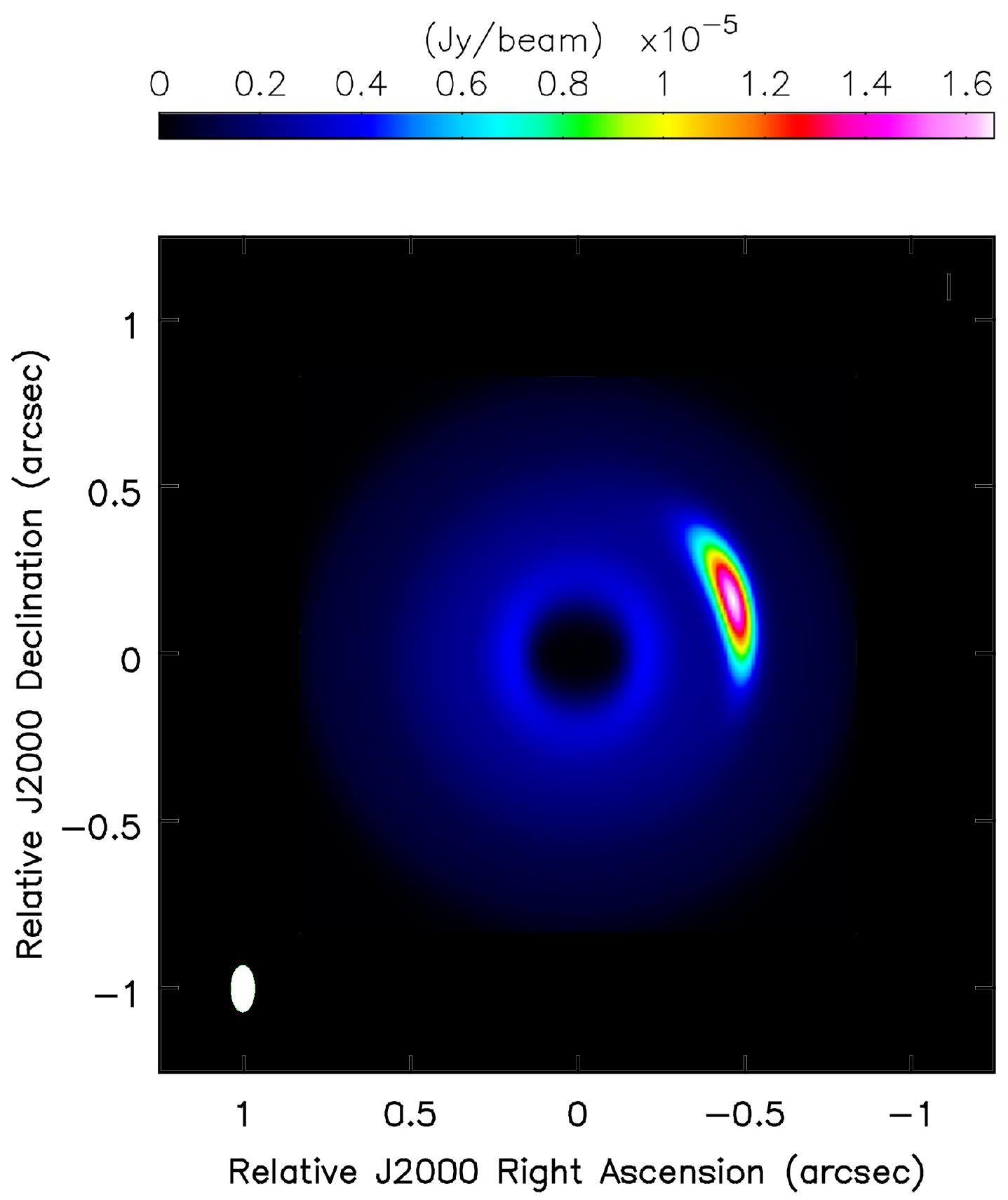}
\caption{Upper panels: Simulated ALMA images at 1 mm (left) and 3 mm (right). Lower panels: Simulated VLA images at 7 mm (left) and 1 cm (right). The beam is shown in the left bottom corner of each image.}
\label{fig:ALMA_VLA_simulations}
\end{figure*}

Figure \ref{fig:ALMA_VLA_simulations} shows the ALMA simulated maps in the upper panels (left: 1 mm, right: 3 mm). 
Lower panels show the VLA maps (left: 7 mm, right: 1 cm). The vortex emission starts to show at 3 mm and becomes very evident at larger wavelengths, in the VLA images.

\begin{figure*}[!t]
\centering
\includegraphics[scale=0.45]{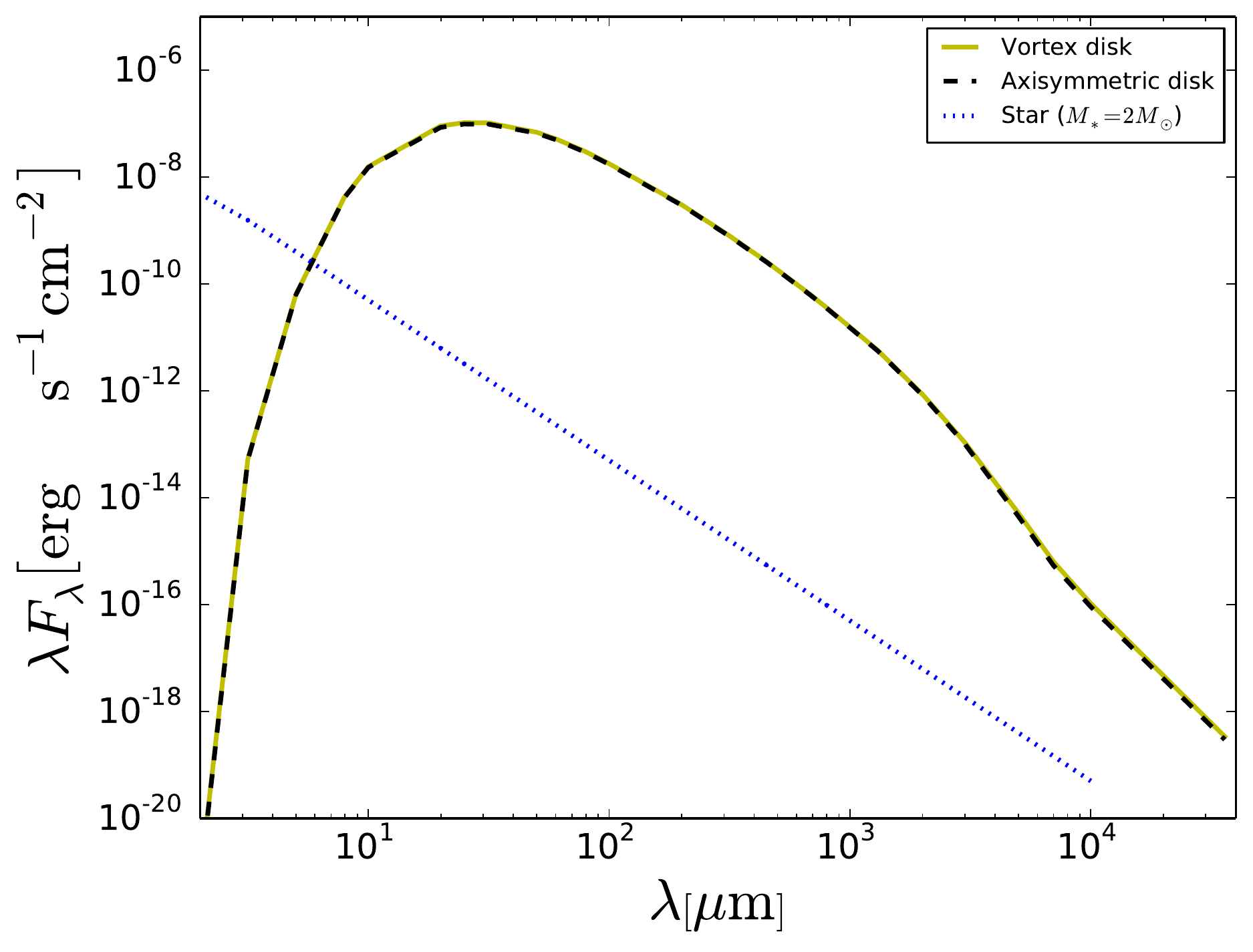} \includegraphics[scale=0.45]{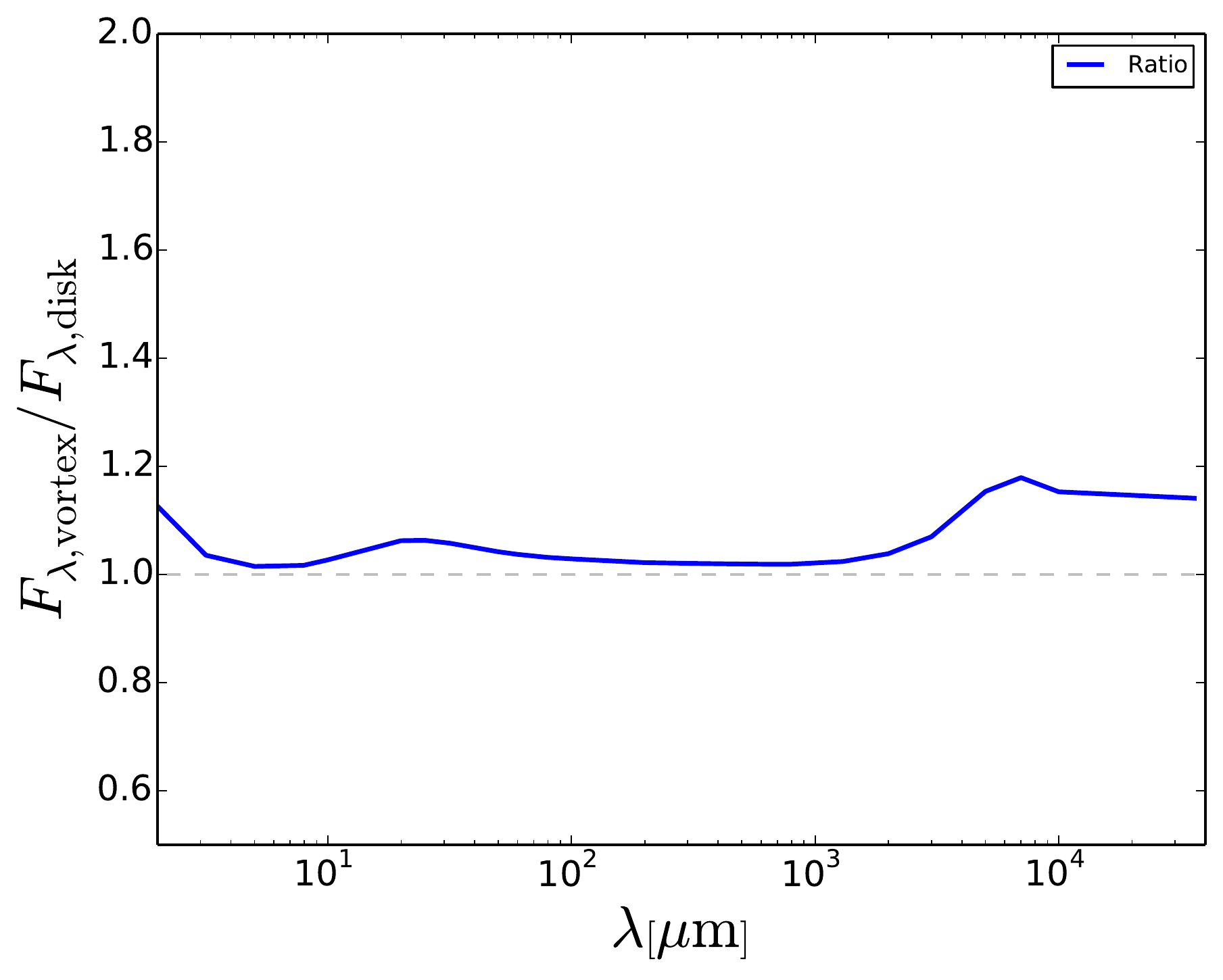}
\caption{Left panel: SED of the vortex disk (yellow solid line), an axisymmetric disk (black dashed line), and a $2 M_{\odot}$ star (blue dotted line). Right panel: Ratio between the flux of the vortex disk and the axisymmetric disk.}
\label{fig:SED}
\end{figure*}

The left panel of Figure \ref{fig:SED} shows the SED of the disk (yellow solid line). We also include the SED of an axisymmetric disk (black dashed line) and the contribution of a 2$M_{\odot}$ star (blue dotted line). Note that the SED is not modified by the emission associated to the vortex, except at mm wavelengths where the disk becomes optically thin. The right panel shows the ratio between the flux from the vortex disk and the axisymmetric disk. The ratio has a maximum value of $1.25$ at $\lambda \sim 7$ mm.  The reason for this behaviour is that 
the numerical simulation of B17 uses an adiabatic equation of state, thus, the disk temperature increases with the surface density; therefore, the flux associated to vortex is larger than the flux from the same region in the axisymmetric disk.

\section{Discussion} \label{sec:discussion}
\subsection{Effect of the Dust Concentration} \label{subsec:dust_distribution}
Strong azimuthal asymmetries at mm wavelengths, as observed, e.g,  in the OpH IRS 48 disk \citep{VanderMarel_2013}, are obtained in the disk models with a concentration of the dust particles around the vortex center, as shown in the Figure \ref{fig:ALMA_VLA_simulations}. The strong disk asymmetry comes from the enhancement of the dust to gas mass ratio due to size segregation inside the vortex, the opacity is increased by a factor of 10 at the vortex center. The smaller slope of the size distribution $p$ within the vortex also increases the opacity at mm wavelengths (see Appendix \ref{sec:opacity}). For example, for $p = 3$ and $\lambda = 7$ mm, the opacity increases by a factor of $\sim 1.8$ compared with the case $p=3.5$. Both effects (higher dust to gas mass ratio and lower slope of the size distribution) are responsible for the increase of the opacity by a factor of 18 at the vortex center.

\begin{figure*}[!t]
\centering
\includegraphics[scale=0.5]{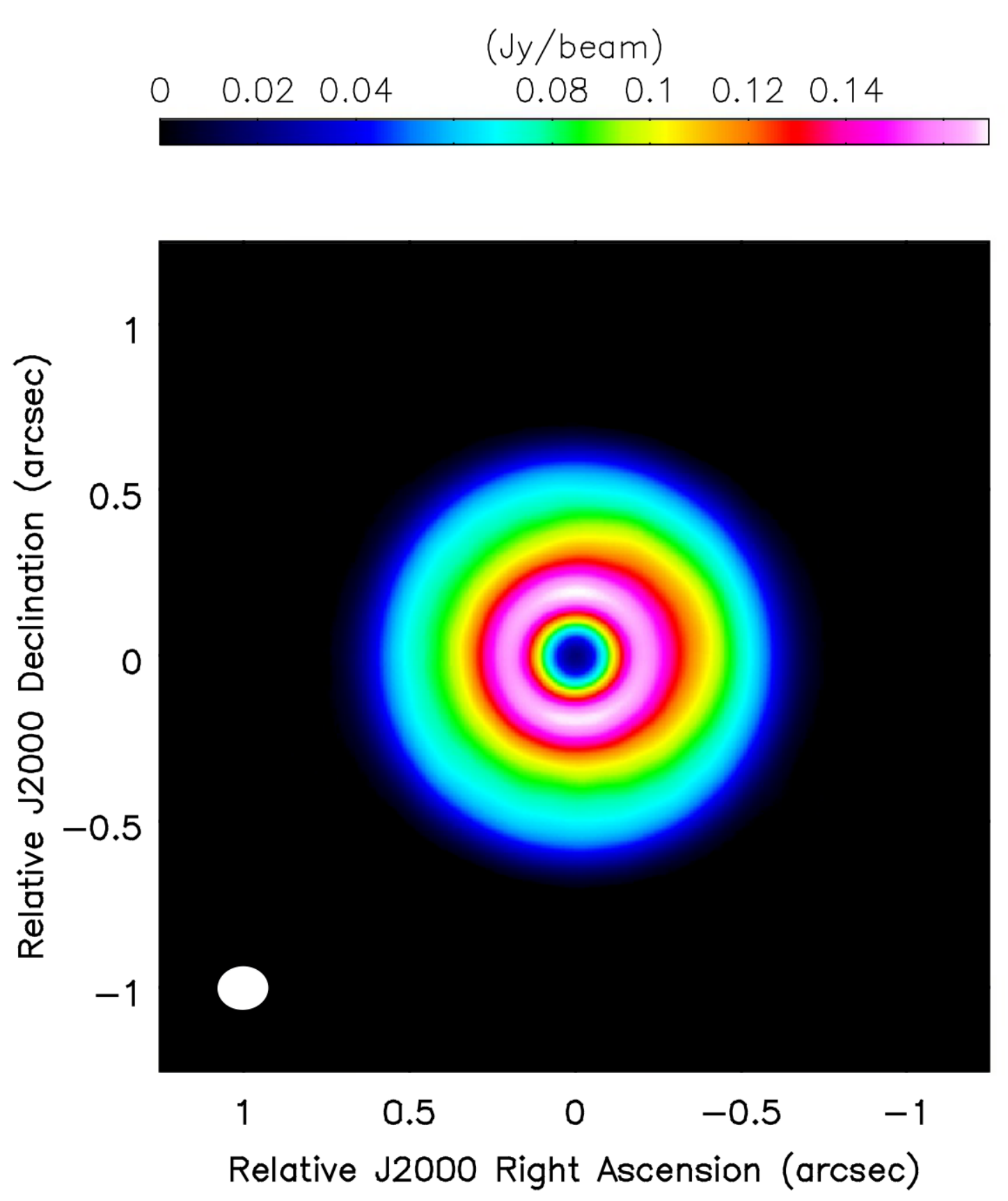}
\includegraphics[scale=0.5]{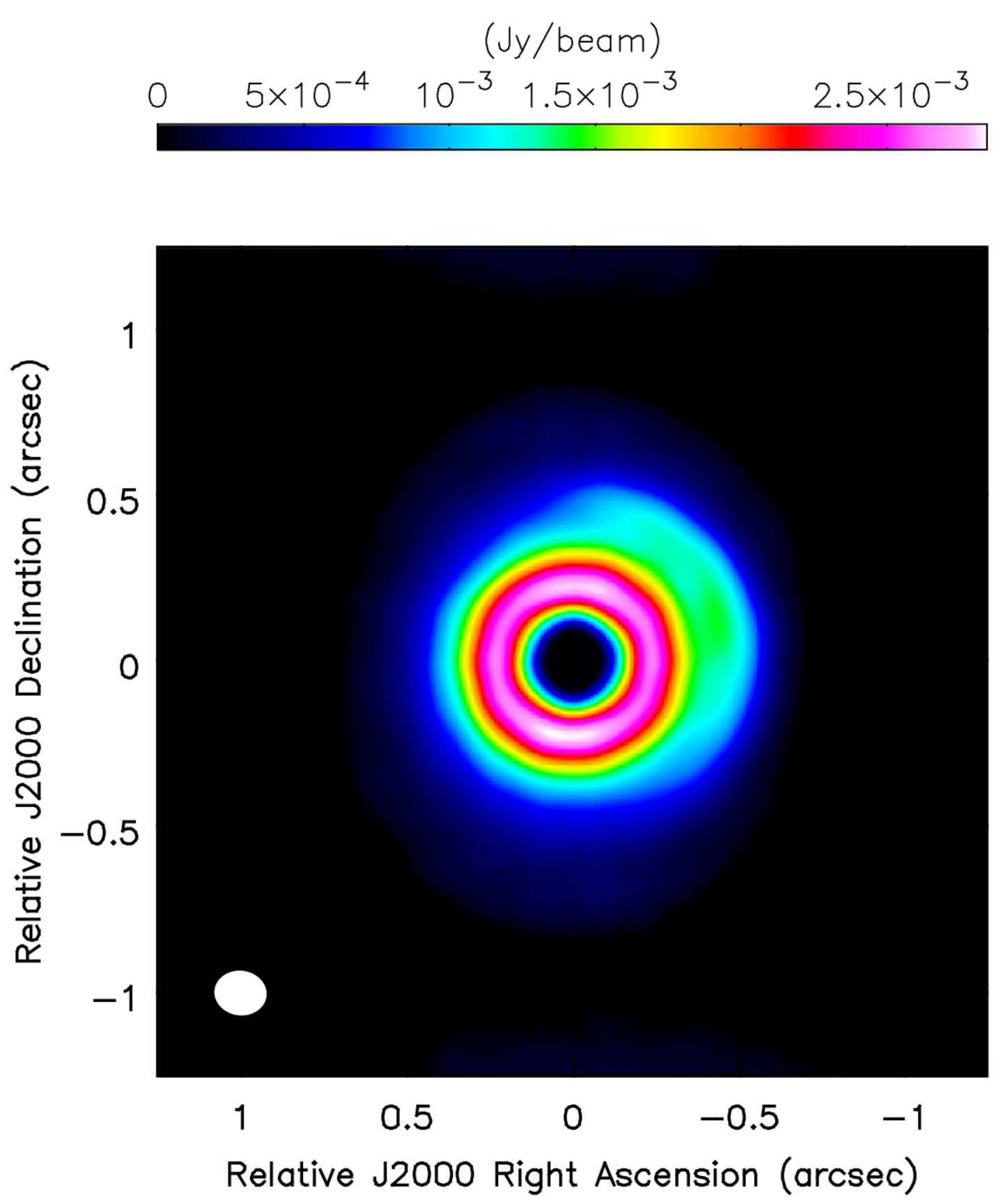}
\includegraphics[scale=0.5]{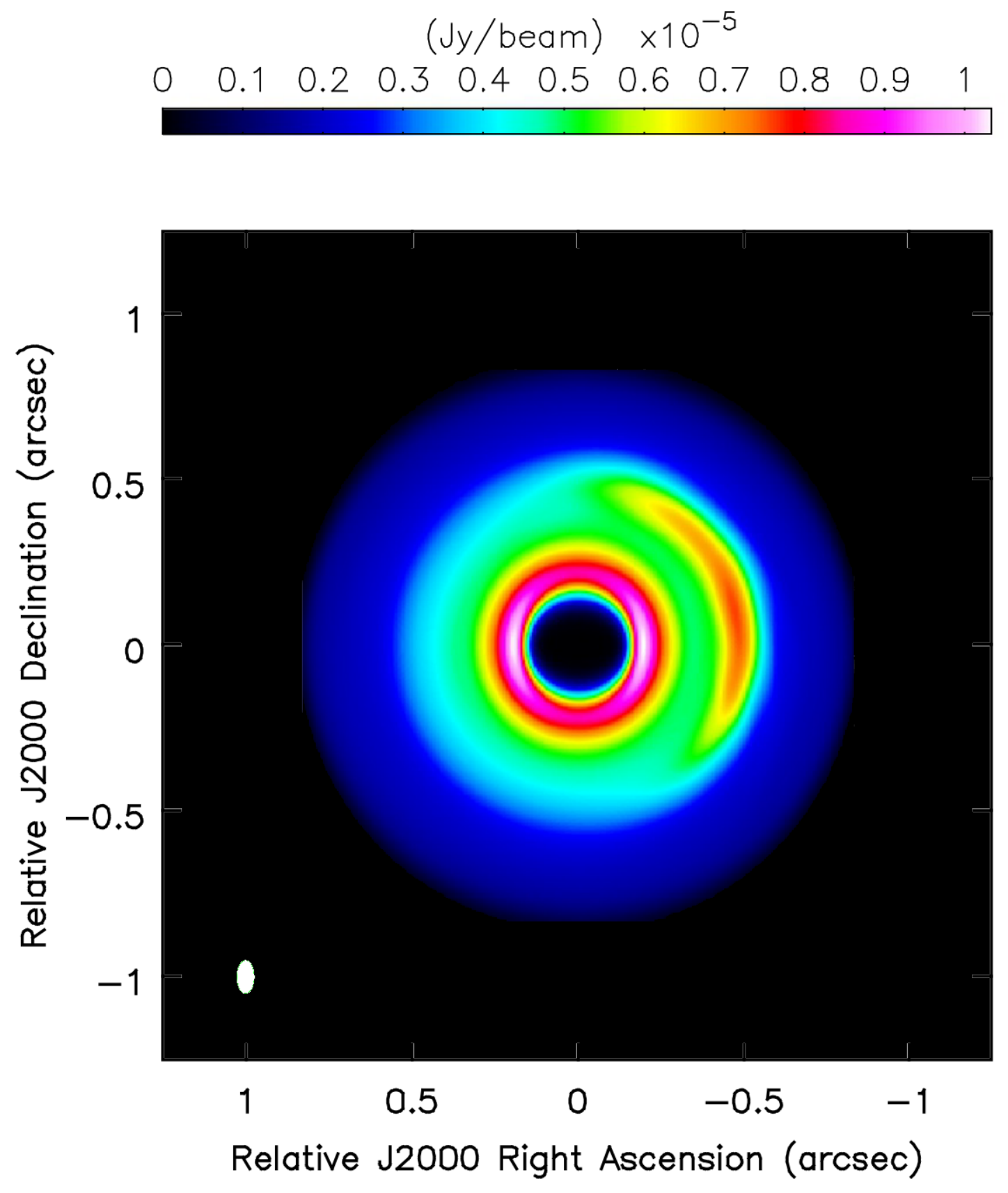}
\includegraphics[scale=0.5]{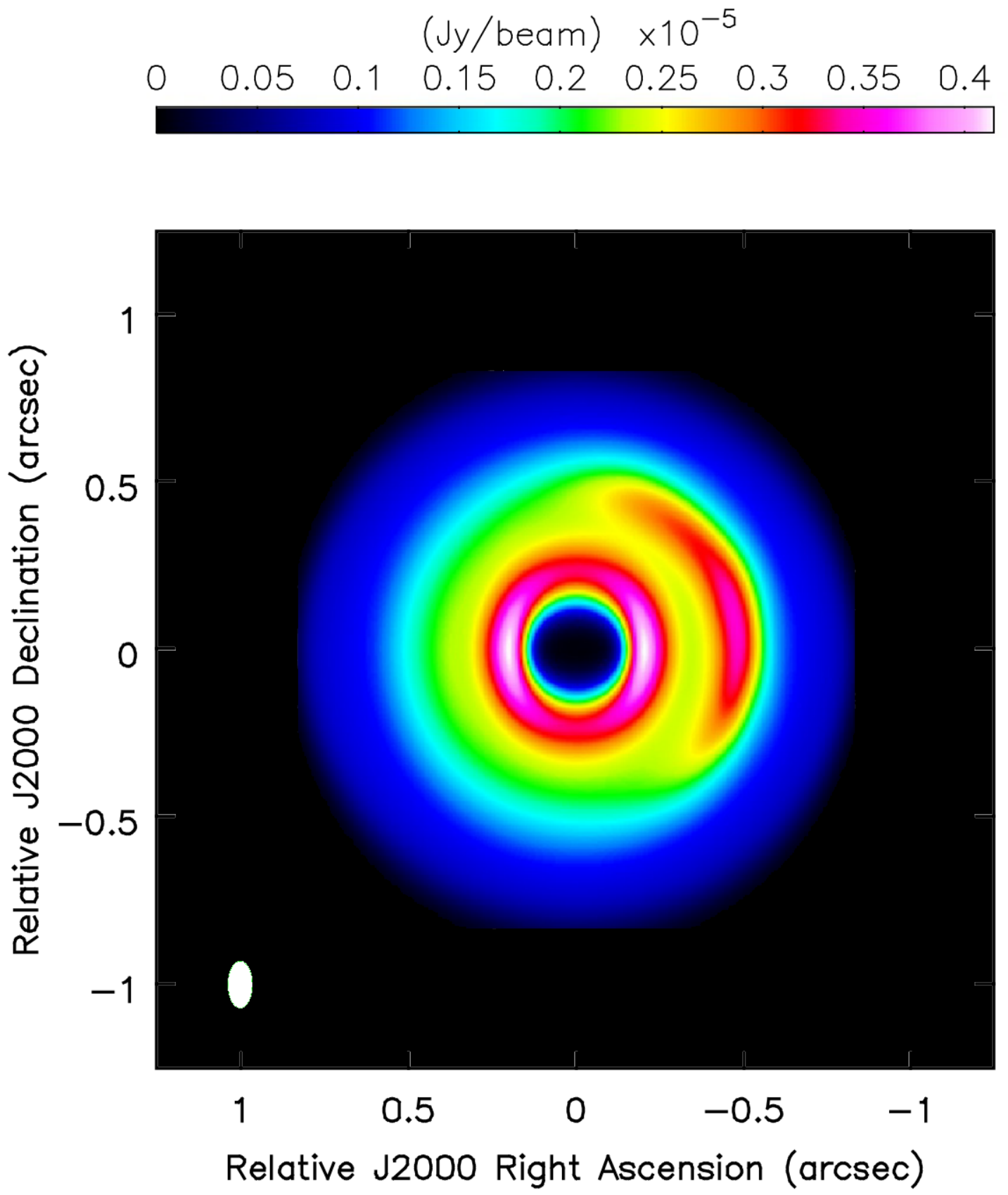}
\caption{Upper panels: Simulated ALMA images at 1 mm (left) and 3 mm (right). Lower panels: Simulated VLA images at 7 mm (left) and 1 cm (right). These models do not include dust segregation inside the vortex (see text). The beam is shown in the left bottom corner of each image.}
\label{fig:ALMA_VLA_simulations_pConst}
\end{figure*}

In order to explore the effect of dust concentration in the vortex, we made maps of the disk model at 1, 3, 7 mm and 1 cm assuming a constant dust to gas mass ratio of $\epsilon = 0.0131$ and a fixed slope of the dust particle size distribution $p =3.5$ throughout  the disk. These maps are shown  in Figure \ref{fig:ALMA_VLA_simulations_pConst} with the same observational parameters described in the subsection \S \ref{subsec:images}. 

The ALMA maps at 1 and 3 mm  do not change much from those in Figure \ref{fig:ALMA_VLA_simulations} where segregation is taken into account. However, there is a dramatic change for the VLA observations at 7 mm and 1 cm. If the dust segregation is not included, the high contrast between the vortex and the disk vanishes. 

\begin{figure*}[!t]
\centering
\includegraphics[scale=0.35]{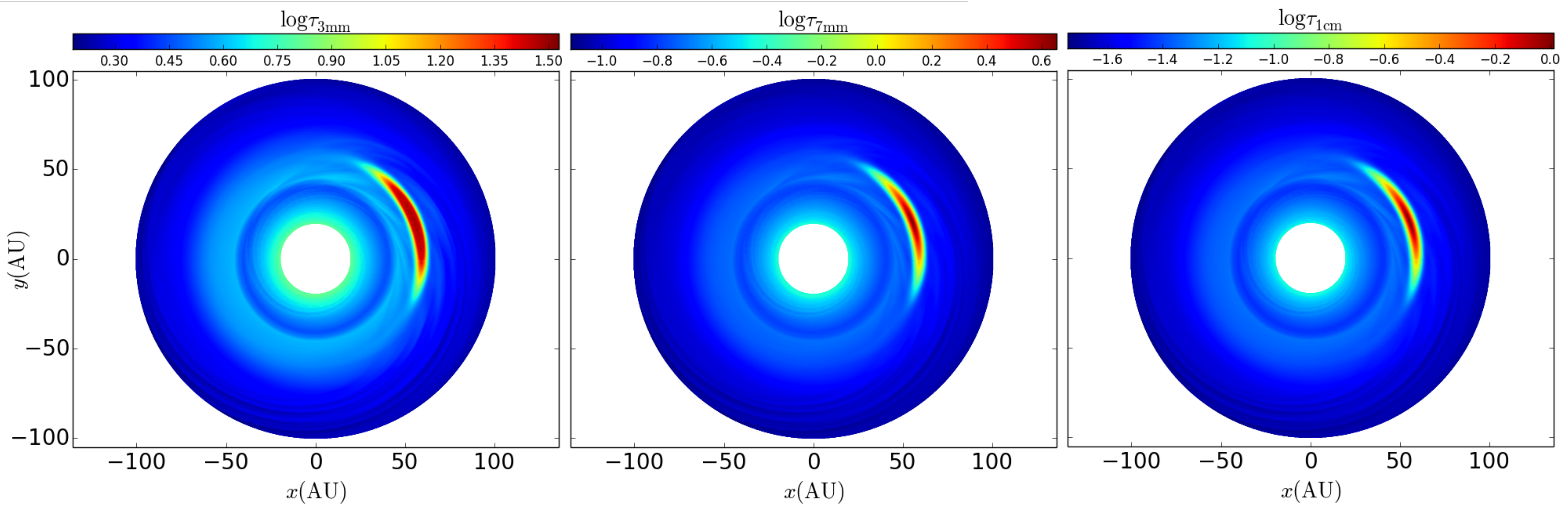}
\caption{Logarithm of the optical depth of the disk at 3 mm (left), 7 mm (middle), and 1 cm (right). }
\label{fig:optical_depth}
\end{figure*}

In the model with dust segregation, the vortex dominates the emission at long wavelengths. For those wavelengths the disk becomes optically thin while the vortex remains optically thick due to the increase of the dust opacity in this region.  
Figure \ref{fig:optical_depth} shows the logarithm of the optical depth at 3, 7 mm and 1 cm.
At 3 mm, all the disk is optically thick ($\log \tau_{3 \rm mm} > 0$). At 7 mm, the vortex is optically thick, but the rest of the disk becomes optically thin ($\log \tau_{7 \rm mm} <0$). At 1 cm, the vortex still remains optically thick, while the background disk is optically thin. Therefore, dust segregation is a crucial ingredient to produce strong azimuthal asymmetries in the dust thermal emission at 7 mm and 1 cm. Note, however, that jets and photoevaporated disk winds will also contribute to the 1 cm emission. Thus, high resolution images are necessary to distinguish the vortex at this wavelength (e.g. \cite{Macias_2016}).

\subsection{Absorption and Scattering} \label{subsec:absorption}
The contribution of the scattering coefficient to the total opacity is important at mm wavelengths, where the albedo is close to 1.  In particular, for large grains with $a_{\rm max}= 1$ mm, $\sigma_\nu > \kappa_\nu$ for mm wavelengths; thus, scattering increases the total opacity and the albedo is large $\omega_{\nu} > 0.8$ (see Figure \ref{fig:opacities_app_2}).

In this section we produce synthetic images without scattering to compare them with the results of the previous section \S \ref{subsec:images}, i.e., we set  $\sigma_{\nu} = 0$, $\omega_{\nu} = 0$ in the eqs. (\ref{eq:opacity}), (\ref{eq:source_function}), and (\ref{eq:zero_moment}). 

\begin{figure*}
\centering
\includegraphics[scale=0.45]{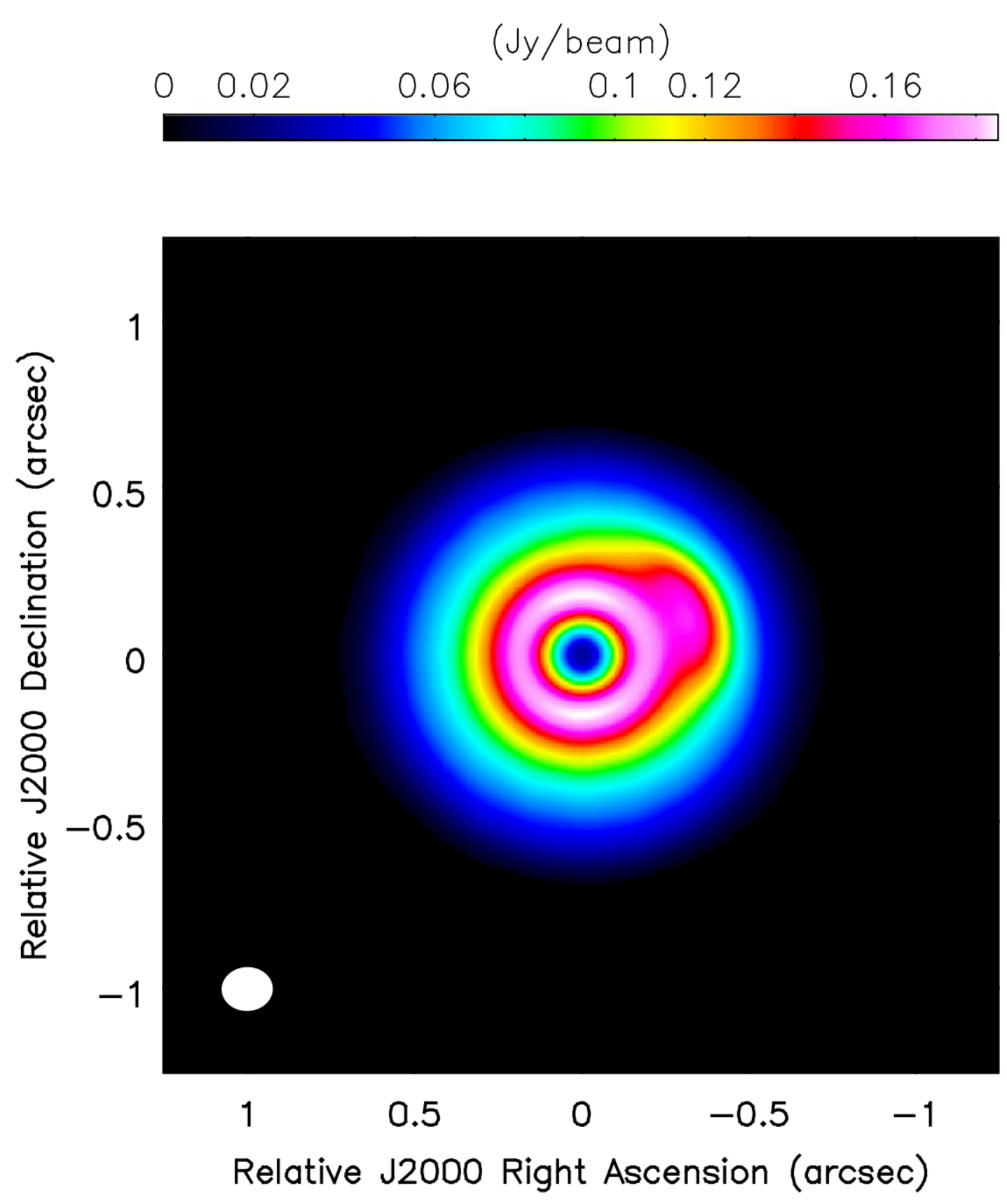}
\includegraphics[scale=0.45]{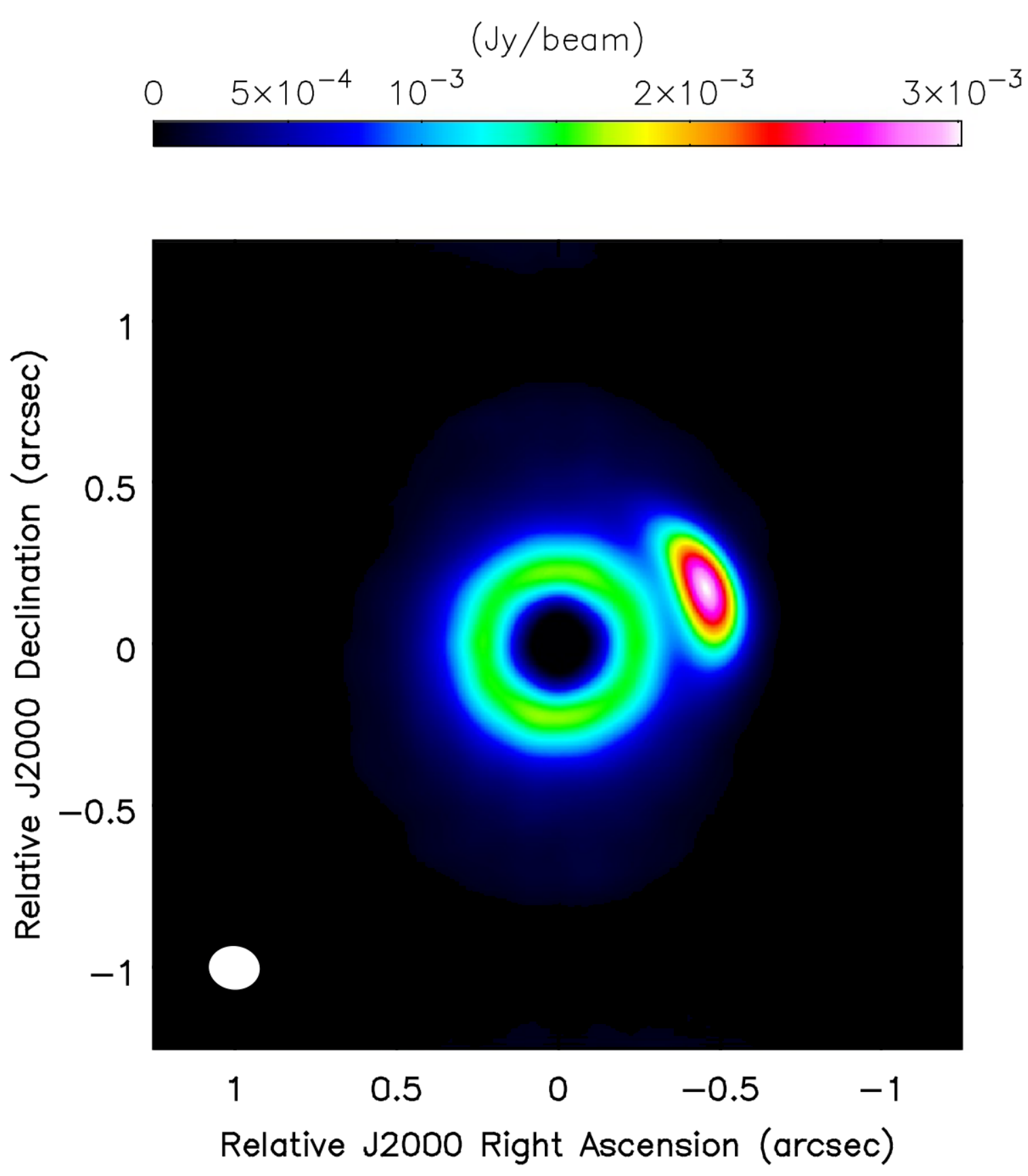}
\includegraphics[scale=0.45]{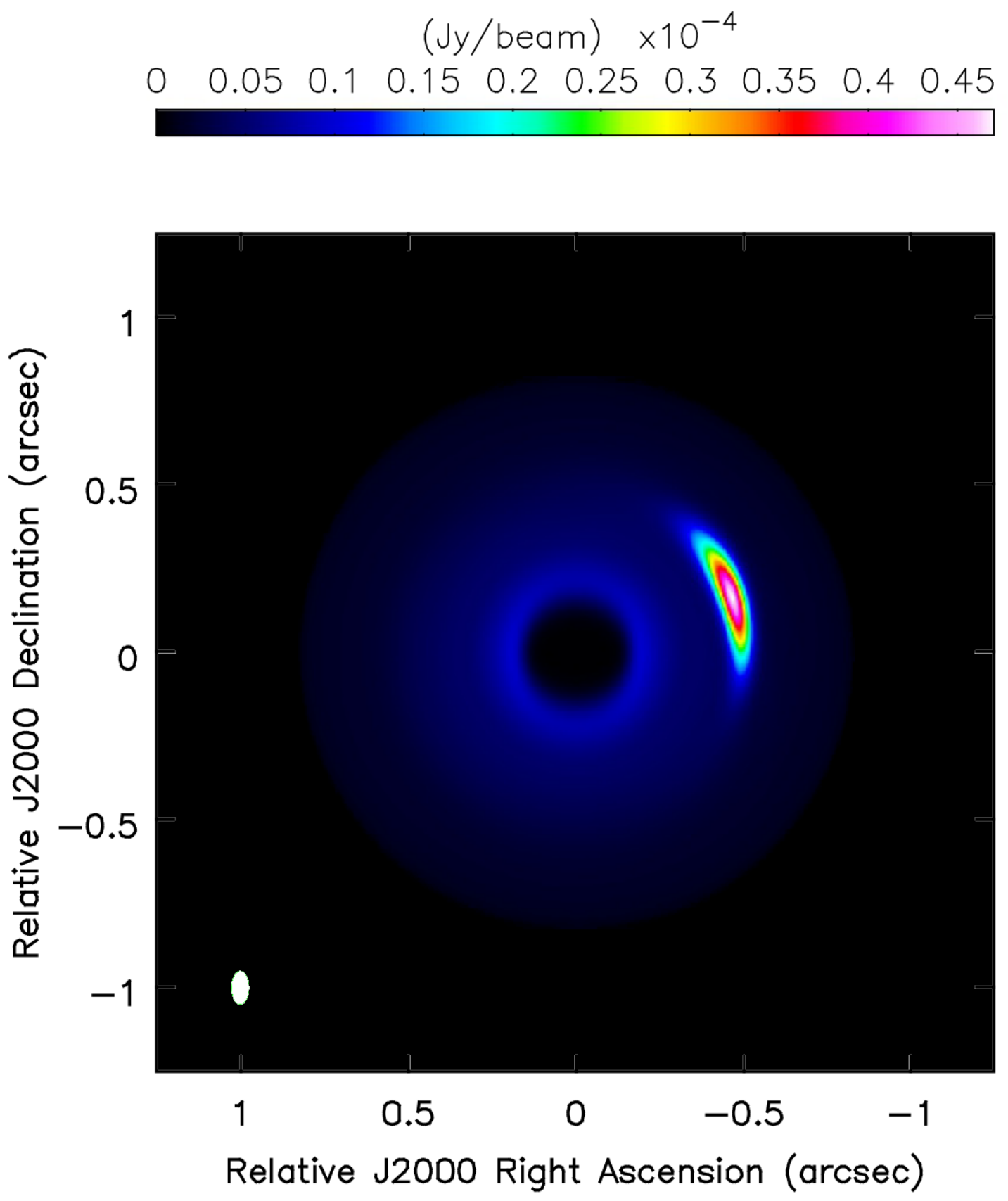}
\includegraphics[scale=0.45]{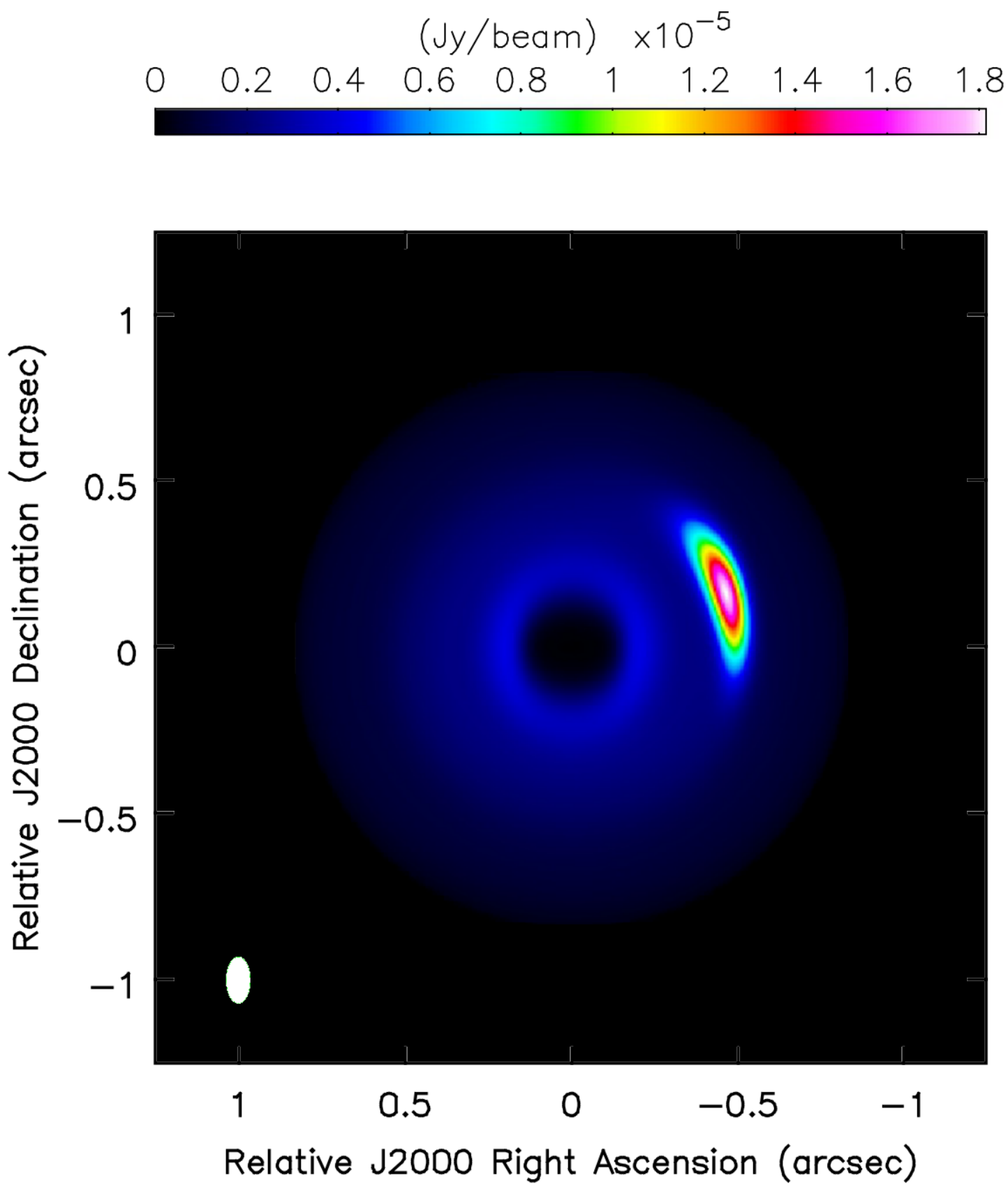}
\caption{Upper panels: Simulated ALMA images at 1 mm (left) and 3  mm(right). Lower panels: Simulated VLA images at 7 mm (left) and 1 cm (right). For these maps the monochromatic scattering coefficient is $\sigma_\nu=0$.}
\label{fig:ALMA_VLA_simulations_abs}
\end{figure*}

Figure \ref{fig:ALMA_VLA_simulations_abs} shows the simulated ALMA maps at 1 and 3 mm and the simulated VLA maps at 7 mm and 1 cm when only the monochromatic absorption mass coefficient is included in the opacity. An important effect of ignoring scattering is that the
emission at all wavelengths is higher than the emission of the maps in Figure \ref{fig:ALMA_VLA_simulations}. 
In addition, the 3 mm map shows a stronger azimuthal asymmetry 
compared with the 3 mm map of  Figure \ref{fig:ALMA_VLA_simulations}.
  This happens because  the optical depth decreases by an order of magnitude 
making the background disk optically thin at 3 mm, 7 mm and 1 cm, while the vortex remains optically thick.

Recently, \cite{Kataoka_2015} found that self-scattering of thermal dust emission in protoplanetary disks produces polarized emission at millimeter wavelengths. The highest degree of polarization occurs for $a_{\rm max} \sim \lambda/2 \pi$, where $\lambda$ is the observing wavelength. Also, in the disk inner region, the direction of the polarization vectors tend to align with the disk minor axis. This property could help to discriminate scattered emission from direct emission from elongated grains aligned perpendicular to the  magnetic field, which is used to determine the magnetic field morphology, e.g, \cite{Rao_2014}. \cite{Yang_2016(2)} studied the relative importance of both scattering and direct emission from magnetically aligned grains as a function of the inclination of the disk in the plane of the sky $i$. They found that the scattering polarization dominates for edge-on ($i \rightarrow 90$ deg) disks. \cite{Kataoka_2016} and \cite{Yang_2016} argue that millimeter emission from the disk around HL Tau can be explained by dust self-scattering, while  in the case of the disk around  NGC 1333 IRAS 4A1, it can be explained as a combination of direct emission by aligned grains and scattering \citep{Yang_2016(2)}. 

\subsection{Dust to Gas Mass Ratio} \label{subsec:dust_to_gas}
\cite{Surville_2016} propose a model to follow the evolution of the dust population in the gas vortex during the linear capture regime that has an invariant
\begin{equation}
I = \epsilon_{l}(1+\beta_{\Omega}/2) + |Ro|,
\end{equation}
where $Ro$ is the Rossby number and $\beta_{\Omega} = d \ln \Omega / d \ln \varpi$ is the slope of the angular velocity of the background disk. This equation is a good approximation for the first hundred rotations of the disk (measured at the vortex orbit). As a function of the time, the Rossby number tends to zero at the center of the vortex \citep{Surville_2016}, so, for a Keplerian disk with $\beta_{\Omega} = -3/2$, the maximum dust to gas mass ratio ($\epsilon_{l, \rm max}$) is only a function of the initial Rossby number and the initial dust to gas mass ratio $\epsilon$
\begin{equation}
\epsilon_{l, \rm max}= 4 |Ro|_{t=0} + \epsilon.
\label{Eq:Surville}
\end{equation}
\cite{Surville_2015} obtained appropriate values of the Rossby number for 300 simulations by varying temperature, surface density and scale height of the disks; they found that most of the vortices have $-0.17 < Ro < -0.11$. With this range of values, the expected dust to gas mass ratio in the vortex center is $\epsilon_{l, \rm max} \sim 0.44 - 0.68$.

The dust to gas mass ratio found at the vortex center in 
section \S \ref{subsec:dust_disk_model}  is $\epsilon_{l,\rm max}/\epsilon \sim 10.4$.
For a standard value $\epsilon = 0.01$, the dust to gas mass ratio predicted in this work is lower by a factor of $\sim 4$ than the value given
by eq. (\ref{Eq:Surville}). It would be important to follow the dust evolution in the vortex disk simulation of B17 used in this work.

\subsection{Maximum grain size} \label{subsec:max_grain_radius}
\begin{figure*}[!t]
\centering
\includegraphics[scale=0.35]{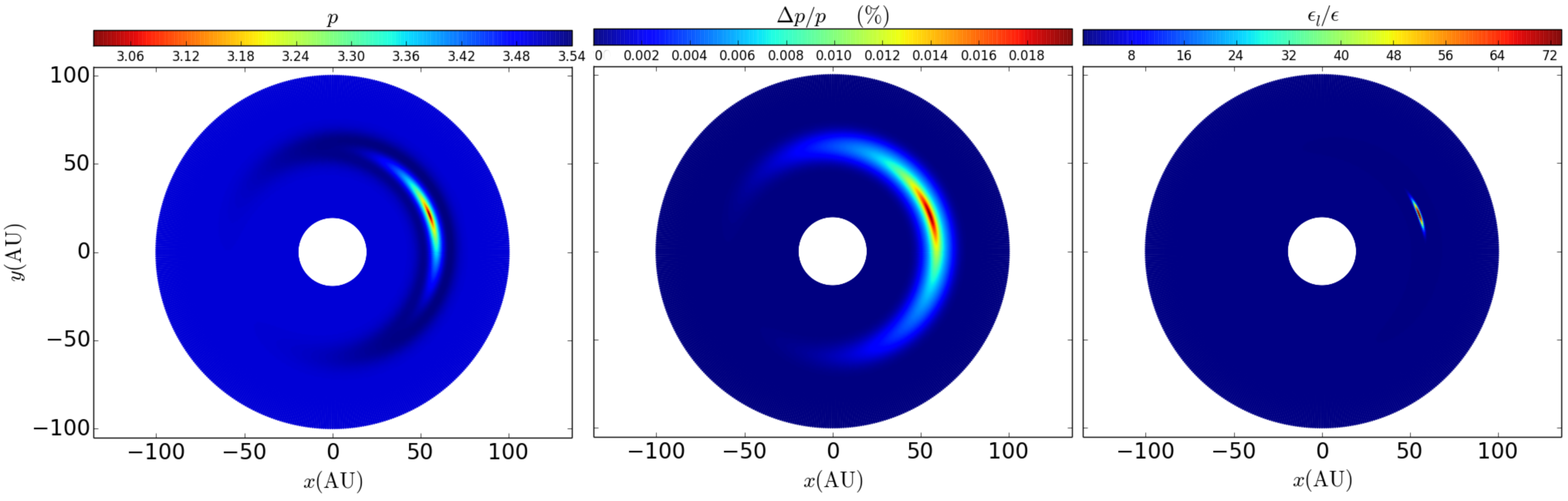}
\caption{Dust particle accumulation within the vortex for $a_{\rm max} = 1$ cm. Left panel: Slope $p$ of the power law fit to the dust particle size distribution $n(a)da \propto a^{-p}da$ (see text). Middle panel: fractional standard deviation  $\Delta p/p$. Right panel: Map of the local dust to gas mass ratio (normalized to $\epsilon$).}
\label{Fig:D2G_1cm}
\end{figure*}

In the previous sections $a_{\rm max}$ was set up to 1 mm; however, observational evidences suggest that the maximum grain size could reach $a_{\rm max} \sim$ 1 cm (e.g., \citealt{Perez_2015}).
Figure \ref{Fig:D2G_1cm} shows the dust properties when $a_{\rm max} = 1$ cm. The slope $p$ does not change significantly compared to the previous model ($a_{\rm max}=1$ mm).  However, the maximum dust to gas mass ratio within the vortex is 7 times larger. This happens because 1 cm grains are more concentrated  toward the vortex center since they have a Stokes number 10 times larger than 1 mm grains (eq. \ref{eq:Dust_surface_density}).

Dust opacity also depends on $a_{\rm max}$ (see Appendix \ref{sec:opacity}). 
For $a_{\rm max} = 1$ cm the opacity  at 1, 3, 7 mm, and 1 cm is modified by a factor of $0.40, 0.55, 5.6, 20$, respectively, compared with the case of $a_{\rm max} = 1$ mm. This means that at 7 mm and 1 cm the vortex emission will dominate the disk emission, but will have a small spatial extent, due to the strong dust concentration. At 1, 3 mm, the increase of the dust to gas mass ratio can overcome the decrease of the opacity; however, the background disk remains optically thick, thus, one expects maps similar to those in Figure \ref{fig:ALMA_VLA_simulations}.

\newpage
\section{Conclusions}\label{sec:conclusions}
Dust emission of vortices in protoplanetary disks is studied using the Lyra-Lin model for the concentration of dust grains towards the vortex center. Their analytic model gives the dust surface density for
 a single particle size.  We have extended this model to the case of a dust size distribution $n(a) \propto a^{-p}$. To extend the model, we assume that  the dust mass for each grain size is conserved in the disk, i.e., the dust grains can change their spatial distribution and segregate in the vortex but cannot change their size.
With this assumption we obtain the dust surface density as a function of the grain radius and the gas surface density. We have applied this model to the disk vortex obtained in the numerical simulations of \cite{Barge_2017}.  Due to dust segregation inside the vortex, the local dust to gas mass ratio $\epsilon_{l}$ increases significantly and 
the slope of the size distribution $p$ decreases.
 We find that dust segregation and the inclusion of the scattering opacity are crucial to describe the azimuthal asymmetry in disk emission at mm wavelengths. Our main results are summarized as follows:
 
1. Dust segregation and concentration inside the vortex significantly increases the local dust to gas mass ratio $\epsilon_{l}$ inside the vortex (Figure \ref{fig:dust_dist_fit_p}, right panel); the maximum $\epsilon_{l, \rm max}/\epsilon \sim 10$ is reached near the vortex center. This high value is due to concentration of sub-mm and mm particles that dominate the mass distribution at the vortex center.
 
2. Dust segregation also changes the slope of the dust particle size distribution. 
In our disk model (eq. \ref{Eq:Dust_particle_size_distribution}), the slope of the dust size distribution is found to be less than the standard value ($p = 3.5$); the slope has a minimum close to the vortex center with $p \sim 3$. This change of slope affects both the absorption and scattering coefficients of the dust population. At mm wavelengths, the ratio between the opacities corresponding to the slopes $p = 3.0$ and $3.5$ is 1.8 times larger (Fig. \ref{fig:opacities_app}).

3. The change of the dust properties due to the segregation of dust particles
in the vortex (dust to gas mass ratio and the slope of the dust size distribution) has an important effect: it
 tends to enhance azimuthal asymmetries in the mm emission  (see the simulated maps  at $\lambda = 7$ mm and 1 cm in Fig. \ref{fig:ALMA_VLA_simulations}). If one only considers the effect of an increase of the dust surface density within the vortex using a constant dust to gas mass ratio $\epsilon$, the vortex region does not dominate the emission of the disk at mm wavelengths (see Fig. \ref{fig:ALMA_VLA_simulations_pConst}). The main difference is that when dust segregation is included, the vortex remains optically thick even at long wavelengths, but the rest of the disk becomes optically thin. Instead, for a uniform dust to gas mass ratio, the vortex is optically thin. 

4. For $a_{\rm max} = 1$ mm, dust scattering affects the disk image at millimeter wavelengths: the scattering mass coefficient increases the opacity of the disk by almost one order of magnitude. If scattering is not included in the monochromatic opacity, the optically thick vortex 
region appears as a dominant structure at $\lambda$ = 3, 7 mm and 1 cm, due to smaller optical depth of the background disk. 

Finally, the dust concentration model we developed extending the work of \cite{Lyra_2013} allows us to predict the disk emission at millimeter wavelengths with high angular resolution observations. It will be interesting to study the dust concentration and its emission in other large scale structures like ring gaps and/or spiral arms.

\textit{Acknowledgments}:  A. S. and S. L. acknowledge support of grants PAPIIT-UNAM IN105815 and CONACyT 238631. Computations were performed using HPC resources from GENCI [TGCC and CINES] (Grant -  x2016047407) and also at LAM on the mib MPI cluster maintained by CESAM. We also thank the anonymous referee for his/her useful comments.

\software{CASA (v 4.7.0) \citep{McMullin_2007}}

\appendix

\section{Opacity} \label{sec:opacity}
Using the code from \cite{Dalessio_2001}, where the Mie theory and the dielectric constants of the dust components are used, we construct the total opacity (absorption + scattering) of the dust particles within the disk as a function of wavelength for the dust abundances described in \cite{Pollack_1994}. The opacity is not only a function of the wavelength, but also a function of the local pressure and temperature. The temperature dependence can be stronger than the pressure dependence when sublimation of different dust species occurs. For example, the ice grains sublimate at a temperature around 150 K for typical disk pressures, which causes a significant decrease of the total opacity.

\begin{figure*}[!t]
\centering
\includegraphics[scale=0.9]{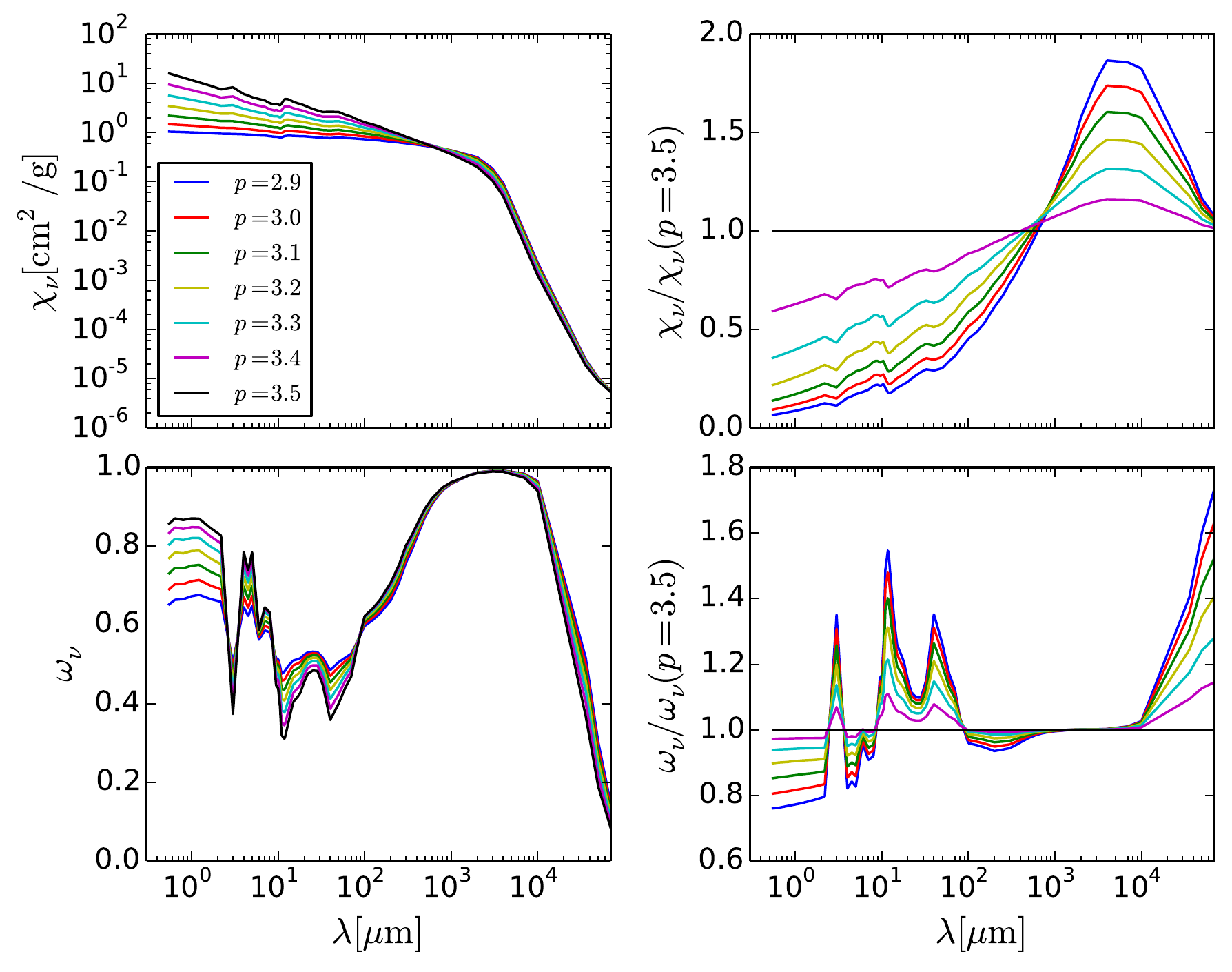}
\caption{Upper panels: dust opacity as a function of wavelength for $T=100$ K. The left panel shows the monochromatic opacity curves ($\chi_{\nu}$) for dust particle size distributions with different slopes $p$.
The right panel shows the monochromatic opacity normalized to the opacity of the standard distribution with slope $p=3.5$.
Lower panels: albedo as a function of wavelength. The left panel shows the monochromatic albedo ($\omega_{\nu}$) for dust particle size distributions with different slopes $p$. The right panel shows the albedo normalized to the albedo of the standard distribution with $p=3.5$. A maximum grain radius of $a_{\rm max} = 1$ mm is assumed in all the cases. The color code is shown in the upper left panel.}
\label{fig:opacities_app}
\end{figure*}

The dust distribution also plays an important role in the magnitude of the opacity. For a given dust particle size distribution $n(a) da \propto a^{-p} da$ that gives the number of particles with sizes between $a$ and $a+da$ per volume unit, the total mean opacity coefficient is

\begin{equation}
\chi_{\nu} =  \frac{\int n(a) a^3 \chi_{\nu}(a) da}{\int n(a) a^3 da}, 
\end{equation}
where $\chi_{\nu}(a)$ is the monochromatic opacity associated to a single particle with dust size $a$. The total opacity is given by  $\chi_\nu = \kappa_\nu + \sigma_\nu$, where $\kappa_\nu$ and $\sigma_\nu$ are the monochromatic absorption  and scattering coefficients, respectively. The albedo is defined as $\omega_{\nu} = \sigma_\nu/\chi_\nu$. The left panels of Figure \ref{fig:opacities_app} shows the opacity $\chi_{\nu}$, and the albedo $\omega_{\nu}$ for different dust size distributions with slopes  $p = 3.5 - 2.9$ in steps of 0.1, and for the dust composition discussed in \S  \ref{subsec:methodology}. In all the cases the opacity units are cm$^2$ per gram of gas. We define a standard opacity given by 
a size distribution with a slope $p = 3.5$. The right panels show the normalized opacity and albedo normalized to the standard case. Note that for short wavelengths, the opacity is lower than the standard value, however, for large wavelengths, the opacity is larger than the standard value. The maximum increase (a factor of 1.8) occurs at $\lambda \approx 7$ mm.\\

\begin{figure*}[!t]
\centering
\includegraphics[scale=0.9]{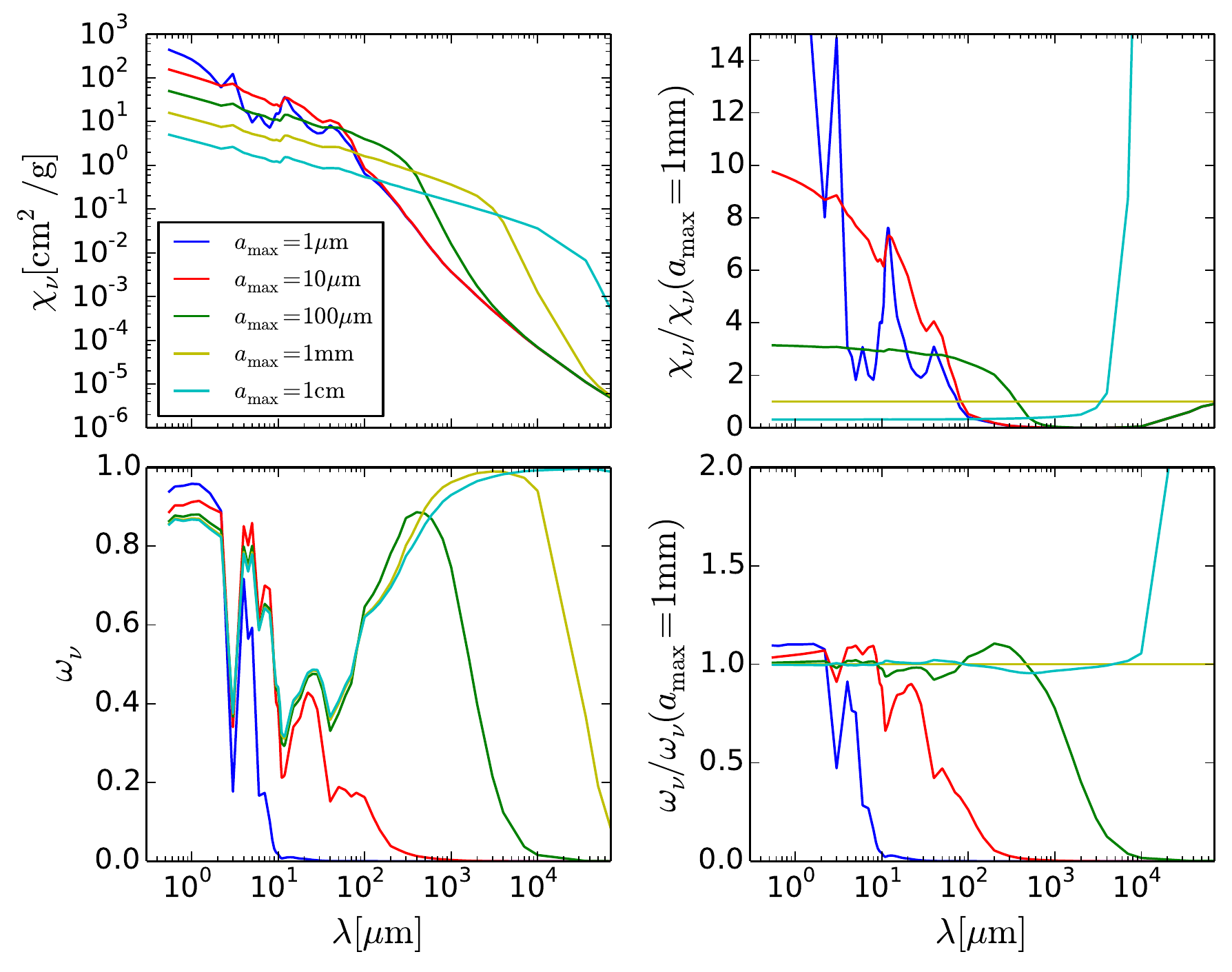}
\caption{Upper panels: dust opacity as a function of wavelength for $T=100$ K. The left panel shows the monochromatic opacity curves ($\chi_{\nu}$) for dust particle size distributions with different maximum grain radii.
The right panel shows the monochromatic opacity normalized to the opacity at $a_{\rm max} = 1$ mm.
Lower panels: albedo as a function of wavelength. The left panel shows the monochromatic albedo ($\omega_{\nu}$) for dust particle size distributions with different maximum grain radii. The right panel shows the albedo normalized to the albedo at $a_{\rm max} = 1$ mm.  A slope of $p = 3.5$ is assumed in all the cases. The color code is shown in the upper left panel.}
\label{fig:opacities_app_2}
\end{figure*}

The opacity and albedo curves are also a function of the maximum grain radius in the dust particle size distribution. The left panels of the Figure (\ref{fig:opacities_app_2}) shows the opacity and albedo as a function of the maximum grain radius. The right panels show the normalized opacity and the albedo normalized with the case $a_{\rm max} = 1$ mm. The albedo becomes important at mm wavelengths for big
grains,   $a_{\rm max}=1$ mm and 1 cm. Although the albedo are similar for these grain sizes,
the opacity for $a_{\rm max}=1$ cm at 1, 3, 7 mm, and 1 cm is a factor of $0.40, 0.55, 5.6, 20$ 
times the opacity for $a_{\rm max}=1$ mm (respectively).

\section{Dust particle size distribution} \label{sec:dust_part_size_dist}
In an isothermal disk, the vertical dust volume density distribution is given by $\rho_d(z) = \rho_{d,0} \exp(-z^2/ 2H_d^2)$, where $H_d$ is the scale height of the dust disk, and $\rho_{d,0}$ is the dust volume density at the midplane. The midplane volume density can be obtained by adding the mass of all the dust particles with material density $\rho_m$ and radius $a$ as

\begin{equation}
\rho_{d,0} = \frac{4\pi \rho_m}{3} \int_{a_{\rm min}} ^{a_{\rm max}} a^3 n(a)da,
\label{eq:dust_volumetric_density_app}
\end{equation}
where $\rho_m$ is the material density, and the dust particle size distribution $n(a) da$. The total dust surface density for the isothermal disk is given by
\begin{equation}
\Sigma_{d} = \sqrt{2\pi} H_d \rho_{d,0},
\end{equation}
and it can also be obtained as the sum of the surface densities of all the dust particles with different sizes 

\begin{equation}
\Sigma_d = \int _{a_{\rm min}} ^{a_{\rm max}} \frac{d \Sigma_{d}(a)}{da} da.
\label{eq:dust_surface_density_app}
\end{equation}

Comparing eqs. (\ref{eq:dust_volumetric_density_app}) and (\ref{eq:dust_surface_density_app}), the dust particle size distribution can be written as
\begin{equation}
n(a)da = c \left[ a^{-3} \frac{d\Sigma_{\mathrm{d}}(a)}{da} \right] da, \quad {\rm with}  \quad c=\frac{3}{ 2^{5/2} \pi^{3/2} H_d \rho_m}.
\label{eq:dust_distribution_sigma}
\end{equation}

For example, when the dust surface density increases with the dust size as $\Sigma_{\mathrm{d}} \propto a ^{1/2}$, one obtains the typical dust particle size distribution with a slope $p = 3.5$.

\listofchanges

\end{document}